\documentclass[11pt]{article}
\usepackage{epsfig} 
\newcommand{\sect}[1]{ \section{#1} \setcounter{equation}{0} }
\newcommand{\partialslash}{\partial \! \! \! /}
\newcommand{\pslash}{p \! \! \! /}

\newcommand{\half}{\mbox{\small{$\frac{1}{2}$}}}

\newcommand{\la}{\langle}
\newcommand{\ra}{\rangle}   
\newcommand{\Nc}{N_{\!c}} 
\newcommand{\Nf}{N_{\!f}} 
\newcommand{\CR}{C_2(R)} 
\newcommand{\CG}{C_2(G)} 
\newcommand{\MSbar}{\overline{\mbox{MS}}} 
\newcommand{\NN}{{\cal N}} 
\newcommand{\OO}{{\cal O}} 
\begin{document}
\title{Three loop renormalization of the $SU(N_c)$ non-abelian Thirring model}
\author{J.F. Bennett \& J.A. Gracey, \\ Theoretical Physics Division, \\ 
Department of Mathematical Sciences, \\ University of Liverpool, \\ 
Peach Street, \\ Liverpool, \\ L69 7ZF, \\ United Kingdom.}
\date{} 
\maketitle
\vspace{5cm}
\noindent
{\bf Abstract.} We renormalize to three loops a version of the Thirring model 
where the fermion fields not only lie in the fundamental representation of a 
non-abelian colour group $SU(\Nc)$ but also depend on the number of flavours,
$\Nf$. The model is not multiplicatively renormalizable in dimensional 
regularization due to the generation of evanescent operators which emerge at 
each loop order. Their effect in the construction of the true wave function, 
mass and coupling constant renormalization constants is handled by considering 
the projection technique to a new order. Having constructed the $\MSbar$  
renormalization group functions we consider other massless independent 
renormalization schemes to ensure that the renormalization is consistent with
the equivalence of the non-abelian Thirring model with other models with a 
four-fermi interaction. One feature to emerge from the computation is the
establishment of the fact that the $SU(\Nf)$ Gross Neveu model is not 
multiplicatively renormalizable in dimensional regularization. An evanescent 
operator arises first at three loops and we determine its associated
renormalization constant explicitly.  

\vspace{-21.5cm} 
\hspace{13.5cm} 
{\bf LTH-456} 

\newpage

\sect{Introduction.} 

Models with a four-fermi interaction have played an important role in exploring
fundamental properties of physical quantum field theories. For example, the two
dimensional $O(N)$ Gross Neveu model, \cite{1}, is similar to QCD in that it is
asymptotically free. Another two dimensional four-fermi theory which shares 
similar properties to the Gross Neveu model is the Thirring model with a 
non-abelian symmetry, \cite{2}. It too is asymptotically free and additionally 
is more closely related to QCD itself. For instance, when the fermion fields of
the non-abelian Thirring model, (NATM), have the same quantum numbers as the 
quarks of QCD, with $\Nf$ flavour dependence in addition to being in the 
fundamental representation of the colour group, it has been argued that it is 
equivalent to QCD in the large $\Nf$ limit, \cite{3}. More precisely, 
Hasenfratz and Hasenfratz demonstrated that by taking the NATM in its 
$d$-dimensional extension then integrating over the leading quark loops which 
dominate when $\Nf$ is large, the usual three and four gluon vertices of QCD 
are recovered as one approaches four dimensions. Moreover, graphs with a quark 
loop and five or more external gluon legs were shown to be irrelevant in the 
four dimensional limit when $\Nf$ was large, \cite{3}. This equivalence can be 
understood better in the context of the critical point renormalization group. 
In effect one has a fixed point equivalence between QCD and the NATM at the 
$d$-dimensional Wilson-Fisher fixed point of QCD which is completely analogous 
to the $d$-dimensional equivalence at the Wilson-Fisher fixed point of $O(N)$ 
$\phi^4$ theory and the two dimensional $O(N)$ $\sigma$ model both treated in 
$d$-dimensions. This underlying $d$-dimensional property of QCD has been 
exploited in the large $\Nf$ expansion to compute $O(1/\Nf)$ corrections to 
various renormalization group functions in the NATM in $d$-dimensions, 
\cite{4,5,6}. More recently, results have also been determined at $O(1/\Nf^2)$,
\cite{7}. As the NATM is equivalent to QCD then the information contained in 
these scheme independent critical exponents also relate to the perturbation 
theory of QCD. The calculational advantage of using the NATM is the absence of 
the graphs with gluon self interactions which substantially reduces the number 
of Feynman diagrams which need to be computed. Given this intriguing connection
between the NATM and QCD it is important to understand the quantum and 
renormalization properties of the former theory in more detail. For instance, 
in QCD the four loop $\beta$-function and mass anomalous dimensions have 
recently been determined in $\MSbar$ in \cite{8,9}. These built on the earlier 
loop calculations of \cite{10,11,12,13}. However, the situation in the NATM is 
unfortunately not as advanced. The model has only been renormalized to two 
loops, \cite{14,15}. Therefore, we have undertaken to renormalize the NATM at 
three loops which is the main topic of this article. Moreover, it is worth 
emphasising that aside from the QCD connection the two dimensional NATM 
deserves study in its own right given its rich structure and connection with 
integrability. Indeed the computation of $S$-matrix elements is closely 
connected with the latter property and their computation in perturbation theory
can only proceed when the model has been rendered finite.  

As will become clear in the discussion, however, performing the renormalization
is a highly non-trivial exercise. Unlike the majority of useful quantum field 
theories the NATM is not {\em multiplicatively} renormalizable in dimensional 
regularization, \cite{15}. Though the original two loop calculation of 
\cite{14} was performed in strictly two dimensions using a cutoff 
regularization, it does not seem appropriate to us for technical reasons to 
develop that regularization for a three loop calculation. Further, dimensional 
regularization is widely used and has been applied to the renormalization of 
four-fermi theories before at three loops in \cite{16,17,18,19}. The lack of 
multiplicative renormalizability is manifested by the generation of extra 
operators in $d$-dimensions which do not have counterterms available from the 
original interactions. However, they differ from the generation of 
non-renormalizable terms in a theory by the fact that they exist only in 
$d$-dimensions and vanish in the limit to two dimensions. Hence, they are 
accorded the name of evanescent operators. As such they affect the extraction 
of the true renormalization group functions which is a problem we address in 
the three loop calculation. This absence of multiplicative renormalizability in
Thirring like models has already been recognized in the work of 
\cite{15,20,21,22} upon which a lot of our calculation is founded. Though in 
\cite{15} the NATM was only discussed at the level of the two loop 
$\beta$-function and anomalous dimension, whilst \cite{21,22} extended the wave
function renormalization to three loops. These papers were important for our 
determination of the $\beta$-function, anomalous dimension and mass anomalous 
dimension at three loops, which is our central aim. Due to these technical 
reasons, which will be explicitly detailed later, we concentrate on presenting 
a comprehensive discussion on the full renormalization of the NATM. Moreover, 
due to the nature of the renormalization we have also found it necessary to 
comprehensively study a variety of mass independent renormalization schemes, 
aside from the widely used $\MSbar$ scheme, to ensure consistency from other 
considerations, \cite{15,21,22}. This is guided very much by the fact that in 
two dimensions Thirring models are equivalent by, for instance Fierz 
rearrangements, to other four-fermi models such as the Gross Neveu model for 
certain values of $\Nf$ and $\Nc$. These properties need to be preserved in the
quantum theory. 

The paper is organised as follows. The basic properties and relevant formalism
is reviewed in section 2 in the context of the known renormalization of
several two dimensional four-fermi models. The fundamental calculational 
details are discussed in section 3 prior to its application first to the 
Gross Neveu model with an $SU(\Nf)$ symmetry in section 4. Here we reproduce
the known three loop $\MSbar$ renormalization group functions prior to  
presenting the detailed renormalizion of the the NATM in the $\MSbar$ scheme in 
sections 5 and 6. We discuss the relation of these results to other 
renormalization schemes in section 7 before providing our conclusions in 
section 8. Various results relevant to the main discussion are provided in an
appendix. 

\sect{Preliminaries.} 

The form of the Lagrangian for the NATM which we will use is 
\begin{equation} 
L^{\mbox{\footnotesize{natm}}} ~=~ i \bar{\psi}^{iI} \partialslash \psi^{iI} 
{}~-~ m \bar{\psi}^{iI} \psi^{iI} ~+~ \frac{g}{2} \left( \bar{\psi}^{iI} 
\gamma^\mu T^a_{IJ} \psi^{iJ} \right)^2 
\label{natmlag} 
\end{equation}  
where we take Dirac fermions $\psi^{iI}$ with an $SU(\Nf)$ internal symmetry,
$1$ $\leq$ $i$ $\leq$ $\Nf$, living in the fundamental representation of the
colour group $SU(\Nc)$, $1$ $\leq$ $I$ $\leq$ $\Nc$. The generators of 
$SU(\Nc)$ are $T^a$ with $1$ $\leq$ $a$ $\leq$ $(\Nc^2-1)$ satisfying the Lie 
algebra 
\begin{equation} 
[ T^a, T^b ] ~=~ i f^{abc} T^c 
\end{equation} 
where $f^{abc}$ are the usual structure constants. With these definitions the
fermion field is endowed with the same structure as quark fields in QCD. The 
more astute reader will have observed that the sign of the coupling constant, 
$g$, is opposite to the normal convention. One reason for this resides in the 
relation the interaction has with the Gross Neveu model which we discuss 
presently. Thus in its present form (\ref{natmlag}) will not be asymptotically 
free. For the major part of the calculation the issue is one of performing the 
renormalization of the model and we will be able to restore the correct 
convention by the change $g$ $=$ $-$ $\lambda$ at the end. Although the main 
focus will indeed be on the NATM, the calculation will run in parallel with a 
similar calculation in the $SU(\Nf)$ Gross Neveu model, \cite{1}. There are 
various reasons for this. First, each possesses a four-fermi interaction with 
the difference residing in the presence of additional Lorentz and colour group 
structure in the former case. More concretely, its Lagrangian with the {\em 
same} fermion definitions is  
\begin{equation} 
L^{\mbox{\footnotesize{gn}}} ~=~ i \bar{\psi}^{i} \partialslash \psi^{i} ~-~ 
m \bar{\psi}^{i} \psi^{i} ~+~ \frac{g}{2} ( \bar{\psi}^i \psi^i )^2 
\label{gnlag} 
\end{equation}  
where, with this choice of coupling constant sign convention, the model is
asymptotically free. Therefore, the integration rules for the Feynman integrals
are the same in each case which will provide us with a useful cross-check on
calculations in the NATM. In other words, since the $d$-dimensional values of
the Gross Neveu graphs are known to three loops, \cite{18}, ensuring that these
are first reproduced exactly gives us confidence that the integration
routines we use are correct. Second, for various values of the parameters of
each model the theories are equivalent. These and other equivalences for the
most general two dimensional four-fermi interaction have been summarized in
\cite{15}. Here we recall two which are relevant for this article. First, the 
abelian limit of (\ref{natmlag}) gives the abelian Thirring model, (ABTM), 
which has the Lagrangian  
\begin{equation} 
L^{\mbox{\footnotesize{abtm}}} ~=~ i \bar{\psi}^{i} \partialslash \psi^{i} ~-~ 
m \bar{\psi}^{i} \psi^{i} ~+~ \frac{g}{2} ( \bar{\psi}^{i} \gamma^\mu 
\psi^{i} )^2 ~.  
\label{abtmlag} 
\end{equation}  
When $\Nf$ $=$ $1$ one can show by Fierz rearrangement that 
\begin{equation} 
(\bar{\psi} \gamma^\mu \psi)^2 ~=~ -~ 2(\bar{\psi} \psi)^2 ~.  
\end{equation} 
Therefore, the $\Nf$ $=$ $1$ abelian Thirring model is classically equivalent
to the $\Nf$ $=$ $1$ Gross Neveu model. Assuming that this is also valid in
the quantum theory, \cite{15,20}, means that the renormalization group
functions have to be the same for $\Nf$ $=$ $1$. Moreover, ensuring that this
is the case is tantamount to preserving the Ward identities which lurk within
the quantum field theory. A second equivalence exists between the $\Nf$ $=$ $3$
Gross Neveu model and the $\Nf$ $=$ $4$ non-abelian Thirring model which was
discussed in \cite{15}. Again we will assume that this is a property of the 
renormalization group functions.  

Indeed in this context it is worth recording the renormalization group 
functions as they stand prior to this work for each model. Although we have 
defined our model as having an $SU(\Nc)$ colour group, the $2$-loop 
$\beta$-function and anomalous dimensions are available for the model with a 
classical Lie group, \cite{2,14}. In our notation, these are  
\begin{equation} 
\beta(g) ~=~ \frac{\CG g^2}{2\pi} ~+~ \frac{T(R)\Nf\CG g^3}{2\pi^2} ~+~ O(g^4)
\label{nabet2} 
\end{equation} 
and 
\begin{equation} 
\gamma(g) ~=~ -~ \frac{C_2(R)T(R)\Nf g^2}{2\pi^2} ~+~ O(g^3)  
\label{nagam2} 
\end{equation} 
where we recall that to this order the $\beta$-function is scheme independent
and the anomalous dimension, $\gamma(g)$, is in the $\MSbar$ scheme. For 
reasons which we explain later, we will concentrate on the group $SU(\Nc)$
whose Casimirs are  
\begin{equation} 
\CR ~=~ \frac{(\Nc^2 - 1)}{2\Nc} ~~~,~~~ \CG ~=~ \Nc ~~~,~~~ T(R) ~=~ 
\frac{1}{2} ~. 
\label{cas} 
\end{equation} 
However, the abelian limit is defined when the renormalization group functions 
are expressed in terms of the Casimirs of a general classical Lie group, by 
setting  
\begin{equation} 
\CR ~\rightarrow~ 1 ~~~,~~~ \CG ~\rightarrow~ 0 ~~~,~~~ T(R) ~\rightarrow~ 1 ~.
\label{abellim} 
\end{equation} 
(For more insight into the relation between the abelian limit of a non-abelian
theory, see \cite{23}). Hence, the abelian Thirring model renormalization group
functions are quite simply  
\begin{equation} 
\beta(g) ~=~ O(g^4) ~~~,~~~ \gamma(g) ~=~ -~ \frac{\Nf g^2}{2\pi^2} ~+~ 
O(g^3) ~. 
\label{abrgedef} 
\end{equation} 
It is worth noting that there exist several all orders results for these
renormalization group functions, \cite{24}. These are 
\begin{equation} 
\beta(g) ~=~ 0 ~~~,~~~ \gamma(g) ~=~ -~ \frac{\Nf g^2} 
{2\pi^2(1 - \Nf g/\pi)} ~.  
\end{equation} 
A similar formula for the $\beta$-function of the NATM when $\Nf$ $=$ $1$ has 
been deduced in \cite{25}. For the Gross Neveu model the renormalization group
functions have been computed to two loops in \cite{1,16} and to three loops in 
$\MSbar$ in \cite{17,18,19} as  
\begin{eqnarray} 
\beta(g) &=& -~ \frac{(\Nf-1)g^2}{\pi} ~+~ \frac{(\Nf-1)g^3}{2\pi^2} ~+~ 
\frac{(2\Nf-7)(\Nf-1)g^4}{16\pi^3} ~+~ O(g^5) \\ 
\gamma(g) &=& -~ \frac{(2\Nf-1)g^2}{8\pi^2} ~+~ 
\frac{(2\Nf-1)(\Nf-1)g^3}{16\pi^3} ~+~ O(g^4) \\  
\gamma_m(g) &=& -~ \frac{(2\Nf-1)g}{2\pi} ~+~ \frac{(2\Nf-1)g^2}{8\pi^2} ~+~ 
\frac{(4\Nf-3)(2\Nf-1)g^3}{32\pi^3} ~+~ O(g^4)  
\label{gnrge} 
\end{eqnarray} 
for an $SU(\Nf)$ internal symmetry. (We note that the results of \cite{17,18} 
corresponded to a Gross Neveu model with Majorana fermions in an $O(\Nf)$
symmetry group. For an even value of this $\Nf$ one can complexify the fermions
to ensure they lie in $SU(\Nf)$.) Setting $\Nc$ $=$ $4$ and $\Nf$ $=$ $1$ in 
(\ref{nabet2}) and (\ref{nagam2}), we find 
\begin{equation} 
\beta(g) ~=~ \frac{2g^2}{\pi} ~+~ \frac{g^3}{\pi^2} ~+~ O(g^4) ~~~,~~~ 
\gamma(g) ~=~ -~ \frac{15g^2}{32\pi^2} ~+~ O(g^3) 
\end{equation}  
and $\Nf$ $=$ $3$ in (2.13), we have to two loops 
\begin{equation} 
\beta(g) ~=~ -~ \frac{2g^2}{\pi} ~+~ \frac{g^3}{\pi^2} ~+~ O(g^4) ~~~,~~~ 
\gamma(g) ~=~ -~ \frac{5g^2}{8\pi^2} ~+~ O(g^3) 
\end{equation}  
which makes explicit the earlier equivalence after accounting for the relative
minus sign in the coupling constant definitions. The apparent non-equality 
between both expressions for $\gamma(g)$ is accounted for by the realisation
that we are comparing the anomalous dimensions in models with four and three 
fields respectively. 

It is worth commenting on the role of the mass term in each of (\ref{natmlag})
and (\ref{gnlag}). Strictly its presence breaks the continuous chiral symmetry 
possessed by each model, \cite{1}. Though in the case of the Gross Neveu model,
which is the same as a model with an $O(2\Nf)$ symmetry, it has a discrete
chiral invariance. Therefore, the equivalences we discussed above will only be
assumed to hold in the massless case. However, if one is considering 
perturbative calculations in this instance one inevitably will encounter
infrared divergences. These can be regularized by using a modified or 
effective fermion propagator as discussed in \cite{15} which is 
\begin{equation} 
\frac{i \pslash}{(p^2 - m^2)} ~.  
\label{masslessprop} 
\end{equation} 
In this case $m$ is not a parameter of the original theory and does not enter
the corresponding renormalization group equation. Indeed the origin of $m$ in
the denominator is not from the Lagrangian but put by hand into the Feynman
integrals which result from a purely massless propagator. For technical reasons
concerned with checking the construction of the renormalization group functions
in a consistent way for various schemes in this case, we will need the full 
renormalization group equation for the massive model. Therefore, we have 
performed the renormalization of each model in two separate cases. One which we
refer to as the massless NATM, where the propagator (\ref{masslessprop}) is 
used and another where the propagator is 
\begin{equation} 
\frac{i (\pslash + m)}{(p^2 - m^2)} 
\label{massiveprop} 
\end{equation} 
which we will refer to as the massive NATM. In this latter case $m$ is treated 
as a parameter of the theory since (\ref{massiveprop}) follows from 
(\ref{natmlag}) and it will appear in the full renormalization group equation. 
The Feynman diagrams will of course be the same for each case. 

The ultraviolet infinities will be regularized by using dimensional 
regularization. In this approach the dimension of spacetime is taken to be 
$d$ $=$ $2$ $-$ $\epsilon$ where $\epsilon$ is small. Integrals are performed
in $d$-dimensions with the poles in $\epsilon$ absorbed into the appropriate
renormalization constants. However, in this analytic continuation the 
$\gamma$-algebra requires special treatment, \cite{26,27,15,21,22} which we 
briefly recall. In $d$-dimensions the set of products of $\gamma$-matrices 
defined by  
\begin{equation} 
\Gamma_{(n)}^{\mu_1 \mu_2 \ldots \mu_n} ~=~ \gamma^{[\mu_1} \gamma^{\mu_2}
\ldots \gamma^{\mu_n]} 
\label{gamndef} 
\end{equation}  
where $[\ldots]$ denotes total antisymmetry, form a complete basis in the
spinor space, \cite{26}. Clearly this space is infinite dimensional, only 
becoming finite when $d$ is restricted to a positive integer. Therefore, in 
dimensional regularization of theories with fermions one, in principle, ought 
to decompose all $\gamma$-strings into this basis. For theories, such as QCD, 
the nature of the interaction and its ultraviolet structure ensures that the 
higher $\Gamma_{(n)}$ do not arise, \cite{21}. The situation for two 
dimensional four-fermi theories is different in that interactions of the form
$(\bar{\psi}\Gamma_{(n)}\psi)^2$ will be generated in the renormalization.
One consequence of this is that the theory is not multiplicatively 
renormalizable in principle and to extract the correct renormalization group 
functions requires care. We remark that this is a feature of dimensional 
regularization and if one were, for example, to use a cutoff in strictly 
two dimensions, the generation of these extra interactions would not arise.
These are the evanescent operators we referred to earlier. In the context of 
the ABTM their presence has been noted before in \cite{15,21,22} and the most 
general (abelian) scenario discussed in \cite{15,20}. In both instances the 
analysis has been to two loops, though in \cite{22} the consequences for the 
three loop wave function renormalization were also discussed. As an aside we 
remark that whilst the three loop Gross Neveu model renormalization has been 
performed, primarily to deduce the renormalization group functions at this 
order, the issue of whether evanescent operators are generated in this model 
had not been resolved, \cite{18,19}. That they do not at two loops has already 
been established. Though it was noted in \cite{22} that their generation at
three loops only affects the form of the four loop renormalization group
functions. As our goal is the full three loop renormalization of 
(\ref{natmlag}) we need to take account of the subtlety of the evanescent 
operators presence. Therefore, we recall the important features of the earlier 
explicit analysis of \cite{15}. At the outset one can acknowledge the existence
of such operators by including them in the most general possible Lagrangian, 
each with its own coupling. To make this more concrete we will focus for the 
moment on the ABTM before remarking on the NATM later. In this instance, the 
$d$-dimensional Lagrangian (\ref{abtmlag}) now becomes  
\begin{equation} 
L^{\mbox{\footnotesize{abtm}}} ~=~ i \bar{\psi}^{i} \partialslash \psi^{i} ~-~ 
m \bar{\psi}^{i} \psi^{i} ~+~ \frac{1}{2} \sum_{\alpha=0}^\infty g_\alpha 
\left( \bar{\psi}^{i} \Gamma^{\mu_1 \cdots \mu_\alpha}_{(\alpha)} \psi^{i} 
\right)^2 
\label{abtmevlag} 
\end{equation}  
where all the fields and parameters are bare and $g_\alpha$ are the couplings
of all the possible four-fermi operators. For notational purposes in general
discussion, we will use Roman letters to refer to only evanescent operators and 
distinguish the coupling of the original operator by $g$. Thus for the ABTM
we take $g_1$ $\equiv$ $g$ and setting $g_k$ $=$ $0$ one recovers the original
Lagrangian, (\ref{abtmlag}). Moreover, when we consider the renormalization
group functions for the model with an infinite number of couplings, we
introduce the convention that if the argument of a function is the original
coupling $g$, then it means that it corresponds to a renormalization group 
function evaluated in the case that $g_k$ $=$ $0$, for all $k$. Greek indices 
will correspond to both original and evanescent couplings and the letters
$\kappa$ and $k$ will be reserved for dummy indices in sums for the respective
cases. 

If, for the moment, we assume that the practical task of computing the Green's
functions to some order has been achieved, then the renormalized parameters
of (\ref{abtmevlag}) can be deduced with respect to some renormalization 
scheme. So, for instance, the renormalized couplings will be a function of
$\epsilon$ and a finite number of other couplings. Therefore, the resulting
$\beta$-function for each coupling will depend on the others. Ordinarily in a 
theory with more than one coupling this is the standard method. However, since
the couplings $g_k$ correspond to operators which do not exist in the limit
$d$ $\rightarrow$ $2$, one might na\"{\i}vely believe that one can recover the 
renormalization group functions of the theory by setting $g_k$ $=$ $0$. The
resulting $\beta$-function of the coupling $g$, however, will not correspond to
the true $\beta$-function of the theory. To understand this, it is best to 
consider the underlying renormalization group equation of the general model, 
(\ref{abtmevlag}). For a renormalized $n$-point Green's function 
$G^{(n)}(p,m,\mu,g_\kappa)$ where $p$ represents the external momenta and 
$\mu$ is a mass scale introduced to ensure the coupling constant remains 
dimensionless in $d$-dimensions, it satisfies, in our conventions,  
\begin{equation} 
\left[ \mu \frac{\partial}{\partial \mu} ~+~ \sum_{\kappa = 0}^\infty 
\tilde{\beta}_{(\kappa)}(g_\alpha) \frac{\partial}{\partial g_\kappa} ~+~ 
\tilde{\gamma}_m(g_\alpha) m \frac{\partial}{\partial m} ~+~ \frac{n}{2} 
\tilde{\gamma}(g_\alpha) \right] G^{(n)}(p,m,\mu,g_\beta) ~=~ 0 
\label{abrge} 
\end{equation}  
where $\tilde{\beta}_{(\kappa)}(g_\alpha)$ are the na\"{\i}ve $\beta$-functions
of the theory and $\tilde{\gamma}(g_\alpha)$ is the na\"{\i}ve anomalous 
dimension. In terms of the respective renormalization constants, $Z_\psi$ and
$Z_{(\kappa)}$, these are defined to be 
\begin{equation} 
\tilde{\beta}_{(\kappa)}(g_\alpha) ~=~ g_\kappa \, \mu \frac{\partial}{\partial 
\mu} \ln Z_{(\kappa)} ~~~,~~~ 
\tilde{\gamma}(g_\alpha) ~=~ \mu \frac{\partial}{\partial \mu} 
\ln Z_\psi ~.  
\end{equation} 
(Our convention for the $\tilde{\gamma}(g_\kappa)$ term of (\ref{abrge}) is 
consistent with the definition of the perturbative functions, 
(\ref{abrgedef}).) By considering the nature of Feynman diagrams, it is simple 
to observe that in the limit $g_k$ $\rightarrow$ $0$, that the na\"{\i}ve 
evanescent $\beta$-functions do not vanish. Therefore there will be a remnant 
contribution in (\ref{abrge}) in this limit, which would contribute to the true
renormalization group functions, which satisfy  
\begin{equation} 
\left[ \mu \frac{\partial}{\partial \mu} ~+~ \beta(g) 
\frac{\partial}{\partial g} ~+~ \gamma_m(g) m\frac{\partial}{\partial m} 
{}~+~ \frac{n}{2} \gamma(g) \right] G^{(n)}(p,m,\mu,g) ~=~ 0 
\label{fullrge} 
\end{equation}  
where $G^{(n)}(p,m,\mu,g)$ now corresponds to a finite $n$-point Green's 
function in strictly two dimensions. In other words, the effect and 
contributions from the evanescent operators in higher loop corrections are not 
properly decoupled in the na\"{\i}ve $\beta$-function, $\tilde{\beta}(g)$. 
Before reviewing how to do this we note our calculational strategy in light of 
the above remarks. Clearly there are two avenues of computation. The first of 
these is to take the most general Lagrangian with evanescent couplings 
truncated at the appropriate order which is dictated by the number of loops the 
renormalization will be performed for. One then carries this out, computing the
na\"{\i}ve renormalization group functions before proceeding to the extraction 
of the true renormalization group functions. However, it is clear that this 
will necessitate a huge amount of calculation since $g_k$ $\neq$ $0$. Since at 
the end of the computation the evanescent couplings will be unimportant, this
provides us with the second approach which is the one we follow. We set $g_k$ 
$=$ $0$ at the outset. This reduces the amount of calculation from the point of
view of Feynman diagrams but we will still be able to deduce the na\"{\i}ve 
functions $\tilde{\beta}_\kappa(g)$ and $\tilde{\gamma}(g)$ which will now only
depend on $g$. The procedure is to begin with the bare operator with a bare 
coupling $g$ and perform the one loop renormalization of the $2$ and $4$-point 
Green's functions. This will generate one or more evanescent operators whose 
coupling will be $O(g^2)$. To perform the two loop renormalization, the 
Lagrangian with these new operators is used with the evanescent operators being
treated as if they were vertices in the original theory. These newly generated 
vertices at this order are included in the subsequent Green's functions and so 
on.  

The correct renormalization group functions in the chosen renormalization
scheme are recovered by applying the projection technique of \cite{28} and 
discussed in the four-fermi context in \cite{15,20}. It is based on the 
observation that the insertion of an evanescent operator in a Green's function 
is not independent of the insertion of the relevant or physical operators in 
the same Green's function. More concretely, if we denote an operator by 
$\OO_\kappa$ and its normal ordered version by $\NN[\OO_\kappa]$, \cite{29}, 
then within the context of a Green's function  
\begin{eqnarray} 
\int d^d x \, \NN [ \OO_k ] &=& \int d^d x \left( \, \rho^{(k)}(g) \NN [ i 
\bar{\psi} \partialslash \psi ~-~ m \bar{\psi}\psi ~+~ 2g \OO_1 ] \right. 
\nonumber \\ 
&& \left. \left. ~~~~~~~~~-~ \rho^{(k)}_m(g) \, \NN[ m \bar{\psi}\psi ] ~+~ 
C^{(k)}(g) \NN [ \OO_1 ] \right) \right|_{ g_i = 0 ~ d = 2 } 
\label{projdef} 
\end{eqnarray}  
where we take $\OO_\kappa$ $=$ $\half (\bar{\psi} \Gamma_{(\kappa)} \psi)^2$ 
and $\rho^{(k)}(g)$, $\rho^{(k)}_m(g)$ and $C^{(k)}(g)$ are the general 
projection functions which will be computed order by order in perturbation 
theory. To do this one inserts an evanescent operator in either a $2$ or 
$4$-point Green's function and calculates it to the appropriate loop order. In 
general this Green's function with the operator insertion will be divergent, 
but one absorbs the regularized infinity into the composite operator
renormalization constant which is available, with respect to the 
renormalization scheme being used. After renormalization one expresses this 
Green's function in two dimensions with $g_k$ $=$ $0$. The result will be 
non-zero. This exercise is repeated for the insertion of each of the operators 
of the right side of (\ref{projdef}) and the finite value after renormalization
is then used to determine the projection functions $\rho^{(k)}(g)$, 
$\rho^{(k)}_m(g)$ and $C^{(k)}(g)$ perturbatively, \cite{15}. The 
practicalities of this procedure will be clarified later when the explicit 
calculations are performed. Therefore, these projection functions are a measure
of the presence of the evanescent operators in the appropriate Green's 
functions. If one now considers the action, $S$, of the theory (\ref{abtmlag}), 
\begin{equation} 
S ~=~ \int d^2 x \, L^{\mbox{\footnotesize{abtm}}} 
\end{equation} 
then it is simple to deduce, \cite{15},  
\begin{eqnarray} 
\mu \frac{\partial S}{\partial \mu} &=& \int d^d x \left( \, 
\tilde{\gamma}(g_\alpha) \NN \left[ i \bar{\psi} \partialslash \psi ~-~ 
m \bar{\psi}\psi ~+~ 2 \sum_{\kappa = 0}^\infty g_\kappa \OO_\kappa \right] 
\right. \nonumber \\ 
&& \left. ~~~~~~~~~-~ \tilde{\gamma}_m(g_\alpha) \, \NN[ m \bar{\psi}\psi ] ~+~ 
\sum_{\kappa = 0}^\infty \tilde{\beta}_{(\kappa)}(g_\alpha) 
\NN [ \OO_\kappa ] \right) ~.  
\end{eqnarray}  
However, substituting for the relation of the evanescent operators to the 
physical ones, results in the simple relations, \cite{15},   
\begin{eqnarray} 
\beta(g) &=& \tilde{\beta}(g) ~+~ \sum_{k=0}^\infty C^{(k)}(g) 
\tilde{\beta}_{(k)}(g) \nonumber \\  
\gamma(g) &=& \tilde{\gamma}(g) ~+~ \sum_{k=0}^\infty \rho^{(k)}(g) 
\tilde{\beta}_{(k)}(g) \nonumber \\  
\gamma_m(g) &=& \tilde{\gamma}_m(g) ~+~ \sum_{k=0}^\infty \rho^{(k)}_m(g) 
\tilde{\beta}_{(k)}(g) 
\label{truerge} 
\end{eqnarray} 
from the realisation that the action of relevant operators obeys 
\begin{equation} 
\mu \frac{\partial S}{\partial \mu} ~=\, \int d^d x \left( \, \gamma(g) 
\NN [ i \bar{\psi} \partialslash \psi ~-~ m \bar{\psi}\psi ~+~ 2 g \OO_1 ] 
~-~ \gamma_m(g) \, \NN[ m \bar{\psi}\psi ] ~+~ \beta(g) \NN [ \OO_1 ] 
\frac{}{} \right) ~.  
\end{equation}  
Following this procedure allows us to properly account for the evanescent 
operator problem. A similar reasoning, which we also found instructive in
understanding the renormalization, arises in the renormalization of the two 
dimensional Wess-Zumino-Witten model, \cite{30,31}. There the evanescent 
operator which is generated occurs in the $2$-point function and its presence 
can be properly accounted for in the construction of the renormalization group 
functions to three loops. The issue of evanescent operators in the context of
renormalization has also been studied in \cite{32,33}. 

Although we have concentrated on the massive models in the above, one can 
easily recover the massless case by setting $m$ $=$ $0$ in the appropriate 
equations.  Further, we have only considered the ABTM, since it is simpler than
the non-abelian case. For the NATM, the procedure is exactly the same. The only
difference being that we have to account for the presence of the generator
$T^a$. For a general colour group there appears to be no simple way of 
proceeding. Indeed a systematic study of group theory for general classical
and exceptional Lie groups in the context of multiloop calculations has been
provided in \cite{34} which illustrates the deep complexity of the problem. 
Instead we follow the Cvitanovic procedure, \cite{35}, and restrict ourselves 
to the colour group $SU(\Nc)$. For this group the generators satisfy 
\begin{equation} 
T^a_{IJ} T^a_{KL} ~=~ \frac{1}{2} \left[ \delta_{IL} \delta_{KJ} ~-~ 
\frac{1}{\Nc} \delta_{IJ} \delta_{KL} \right] 
\label{cvit} 
\end{equation} 
which means that for each $\Gamma_{(n)}$ there are two possible evanescent 
operators. Moreover, it will lead to a substantial simplification in the 
computation where we use (\ref{cvit}) to decompose the interaction of 
(\ref{natmlag}). After completing the calculation at each loop order, we then 
reconstruct the generator products. However, the full colour basis requires the
inclusion of the identity $I \otimes I$ as well as $T^a \otimes T^a$ for 
completeness. Therefore, we take as the generalized Lagrangian  
\begin{equation} 
L ~=~ i \bar{\psi}^{iI} \partialslash \psi^{iI} ~-~ m \bar{\psi}^{iI} \psi^{iI}
{} ~+~ \frac{1}{2} \sum_{\kappa = 0}^\infty \left[ g_{\kappa 0} 
\left( \bar{\psi}^{iI} \Gamma_{(\kappa)} \delta_{IJ} \psi^{iJ} \right)^2 ~+~ 
g_{\kappa 1} \left( \bar{\psi}^{iI} \Gamma_{(\kappa)} T^a_{IJ} \psi^{iJ} 
\right)^2 \right]  
\end{equation}  
where the second subscript in the evanescent coupling counts the number of 
generators. For the NATM, we now reserve $g$ $\equiv$ $g_{11}$ for the coupling
of the physical operator, with the projection functions now denoted by  
$\rho^{(ki)}(g)$, $\rho^{(ki)}_m(g)$ and $C^{(ki)}(g)$ with $i$ $=$ $0$ or 
$1$. Although we have now explicitly specified $SU(\Nc)$ one can reconstruct 
the colour group Casimirs for an arbitrary classical Lie group at the end of 
the renormalization group function calculation by applying the results 
(\ref{cas}) where $T(R)$ always accompanies one power of $\Nf$. One issue which
might arise with this is the appearance of Casimirs other than products of 
(\ref{cas}). From the nature of the Feynman diagrams, it is easy to convince 
oneself that new Casimirs, such as those which appeared at four loops in the 
$\MSbar$ $\beta$-function and quark mass anomalous dimensions in QCD, 
\cite{8,9}, will not occur at three loops in the NATM. 
 
\sect{Calculational details.} 

We now turn to a discussion of the technical aspects of the algorithm used to
carry out the full three loop renormalization of (\ref{natmlag}). We performed
the calculation by a semi-automatic symbolic manipulation method using several
languages and packages. For instance, the Feynman diagrams were generated using
the package {\sc Qgraf}, \cite{36}, and then encoded in the language 
{\sc Form}, \cite{37}, as it was the most favourable package to perform the 
tedious amounts of algebra associated with the $\gamma$-matrices which we now 
discuss in detail. The advantage of using such packages can be fully understood
from the nature of the problem we have already discussed. We are primarily 
interested in the structure of the Gross Neveu and NATM and to a lesser extent 
the ABTM. Our programmes are organised such that the graphs are encoded with a 
general vertex function carrying all the necessary Lorentz, spinor and group 
indices. One then specifies the model by calling the routine which substitutes 
for the vertex peculiar to that model. This degree of flexibility has been 
fundamental in checking the results for the Gross Neveu model which is taken as
the basic reference point. For any topology the evanescent vertex operator can 
also be implemented by using the same encoding for that topology but with the 
basic vertex replaced by the evanescent operator Feynman rule. This ensures 
that no graphs are accidentally omitted. For completeness we record the 
structure of the Feynman rules we have used for the vertices in each model in 
addition to (\ref{masslessprop},\ref{massiveprop}). For the Gross Neveu model 
the basic interaction of (\ref{gnlag}) leads to the Feynman rule 
\begin{equation} 
i g \left[ \delta^{ij} \delta^{kl} \delta_\alpha^{~\beta} 
\delta_\gamma^{~\delta} ~-~ \delta^{il} \delta^{kj} \delta_\alpha^{~\delta} 
\delta_\gamma^{~\beta} \right]  
\end{equation} 
for external fields $\psi^i_\alpha$, $\bar{\psi}^j_\beta$, $\psi^k_\gamma$
and $\bar{\psi}^l_\delta$ labelled anti-clockwise. Whilst the NATM has the 
Feynman rule for the set of fields, $\psi^{iI}_\alpha$, 
$\bar{\psi}^{jJ}_\beta$, $\psi^{kK}_\gamma$ and $\bar{\psi}^{lL}_\delta$, 
\begin{equation} 
i g \left[ \delta^{ij} \delta^{kl} T^a_{JI} T^a_{LK} 
\gamma_{~\alpha}^{\mu ~\beta} \gamma_{\mu \, \gamma}^{~~~\delta} ~-~ 
\delta^{il} \delta^{kj} T^a_{LI} T^a_{JK} \gamma_{~\alpha}^{\mu~\delta} 
\gamma_{\mu \, \gamma}^{~~~\beta} \right]  
\end{equation} 
and for the evanescent vertices  
\begin{equation} 
i g \left[ \delta^{ij} \delta^{kl} \delta_{JI} \delta_{LK} 
\Gamma_{(n) \, \alpha}^{~~~~ ~\beta} \Gamma_{~~~ \gamma}^{(n) ~\delta} ~-~ 
\delta^{il} \delta^{kj} \delta_{LI} \delta_{JK} 
\Gamma_{(n) \, \alpha}^{~~~~ ~\delta} \Gamma_{~~~ \gamma}^{(n) ~\beta} \right]  
\end{equation} 
and 
\begin{equation} 
i g \left[ \delta^{ij} \delta^{kl} T^a_{JI} T^a_{LK} 
\Gamma_{(n) \, \alpha}^{~~~~ ~\beta} \Gamma_{~~~ \gamma}^{(n) ~\delta} ~-~ 
\delta^{il} \delta^{kj} T^a_{LI} T^a_{JK} \Gamma_{(n) \, \alpha}^{~~~~ ~\delta} 
\Gamma_{~~~ \gamma}^{(n) ~\beta} \right] ~.  
\end{equation} 
The abelian limit of the original vertex is 
\begin{equation} 
i g \left[ \delta^{ij} \delta^{kl} \gamma_{~\alpha}^{\mu ~\beta} 
\gamma_{\mu \, \gamma}^{~~~\delta} ~-~ \delta^{il} \delta^{kj} 
\gamma_{~\alpha}^{\mu~\delta} \gamma_{\mu \, \gamma}^{~~~\beta} \right] ~.  
\end{equation} 

To appreciate the initial problems with the $\gamma$-matrices, we recall that
each propagator has one $\gamma$-matrix whilst for the NATM each basic vertex
contributes two more $\gamma$-matrices. Thus for any three loop topology there
are at most fourteen $\gamma$-matrices. For the massive model each topology
will also involve sums of less numbers of $\gamma$-matrices as well. From the
fact that there are four free spinor indices the overall result will be the
tensor product of two $\gamma$-strings. Given the nature of the vertex the
two extremes of this tensor product will take the form $\gamma^{\mu_1} \cdots
\gamma^{\mu_{13}} \otimes \gamma^{\nu_1}$ and $\gamma^{\mu_1} \cdots 
\gamma^{\mu_7} \otimes \gamma^{\nu_1} \cdots \gamma^{\nu_7}$ where the nature 
of the explicit contractions are not important for the present illustration. As
we need to re-express the final value for the integral after integration in 
terms of the $\Gamma_{(n)}$-basis we need to have a systematic way of reducing 
these strings to products of the form  $\gamma^{\mu_1} \cdots \gamma^{\mu_r} 
\otimes \gamma_{\mu_1} \cdots \gamma_{\mu_r}$, $r$ $\leq$ $7$. In the two 
extreme cases there are no free Lorentz indices on the strings. For the former 
it is clear that the string of thirteen $\gamma$-matrices will have six pairs 
of contracted indices. In the latter if there is a pair of contracted indices 
in one string then there will be a pair in the other string also contracted. 
Otherwise the contractions are across the tensor product. In general, rules for
the treatment of the $\Gamma_{(n)}$-basis have been developed, \cite{27}, and 
then applied to the three loop $2$-point function, \cite{21,22}. We have not 
used the majority of these as they prove too cumbersome to implement 
symbolically. Instead we used a minimal number of their properties in addition 
to the $d$-dimensional Clifford algebra  
\begin{equation} 
\{ \gamma^\mu , \gamma^\nu \} ~=~ 2 \eta^{\mu\nu} ~.  
\label{gamalg} 
\end{equation} 
Our algorithm was based on (\ref{gamalg}) to shuffle contracted $\gamma$'s
together via, for example,  
\begin{equation} 
\gamma^\mu \gamma^{\nu_1} \ldots \gamma^{\nu_n} \gamma_\mu \ldots ~=~  
-~ \gamma^\mu \gamma^{\nu_1} \ldots \gamma_\mu \gamma^{\nu_n} \ldots ~+~  
2\gamma^{\nu_n} \gamma^{\nu_1} \ldots \gamma^{\nu_{n-1}} \ldots  ~~~~. 
\end{equation} 
Then contracted pairs were removed with the usual simple identities 
\begin{eqnarray} 
\gamma^\mu \gamma^\nu \gamma_\mu &=& - ~ (d-2) \gamma^\nu \\  
\gamma^\mu \gamma^\nu \gamma^\sigma \gamma_\mu &=& (d-4) \gamma^\nu 
\gamma^\sigma ~+~ 4\eta^{\mu\nu} \\  
\gamma^\mu \gamma^\nu \gamma^\sigma \gamma^\rho \gamma_\mu &=& -~ (d-6) 
\gamma^\nu \gamma^\sigma \gamma^\rho ~-~ 4\eta^{\nu\sigma} \gamma^\rho ~+~
4 \eta^{\nu\rho} \gamma^\sigma ~-~ 4 \eta^{\sigma\rho} \gamma^\nu ~.  
\label{gamprod} 
\end{eqnarray}  
This process was repeated until one was left with tensorial $\gamma$-strings
which only had Lorentz contractions across the tensor product. At this point
we decomposed each $\gamma$-string into its $\Gamma_{(n)}$ basis, which 
requires inverting the simple definition, (\ref{gamndef}). This is achieved as
follows by using the lemmas, \cite{15,27},  
\begin{eqnarray} 
\Gamma^{\mu_1 \ldots \mu_n}_{(n)} \gamma^\nu &=& 
\Gamma^{\mu_1 \ldots \mu_n \nu}_{(n+1)} ~+~ \sum_{r=1}^n (-1)^{n-r} \,  
\eta^{\mu_r \nu} \, \Gamma^{\mu_1 \ldots \mu_{r-1} \mu_{r+1} \ldots 
\mu_n}_{(n-1)} \\  
\gamma^\nu \Gamma^{\mu_1 \ldots \mu_n}_{(n)} &=& 
\Gamma^{\nu \mu_1 \ldots \mu_n}_{(n+1)} ~+~ \sum_{r=1}^n (-1)^{r-1} \,  
\eta^{\mu_r \nu} \, \Gamma^{\mu_1 \ldots \mu_{r-1} \mu_{r+1} \ldots 
\mu_n}_{(n-1)} ~. 
\label{lemm} 
\end{eqnarray} 
Beginning with the product 
\begin{equation} 
\Gamma^\mu_{(1)} \gamma^\nu 
\end{equation} 
we evaluate it in two ways. One way, of course, is simple in that 
$\Gamma^\mu_{(1)}$ $\equiv$ $\gamma^\mu$ and thus the expression is a string of
two ordinary $\gamma$-matrices, $\gamma^\mu \gamma^\nu$. However, using the
lemma once this is equivalent to  
\begin{equation} 
\Gamma^{\mu\nu}_{(2)} ~+~ \eta^{\mu\nu} \Gamma_{(0)} 
\label{gamn2} 
\end{equation} 
where $\Gamma_{(0)}$ carries only spinor indices. It is easy to observe that
this method is iterative. Beginning now with $\Gamma^{\mu}_{(1)}\gamma^\nu 
\gamma^\sigma$ and multiplying (\ref{gamn2}) by $\gamma^\sigma$ and applying 
(\ref{lemm}) gives the relation between $\Gamma^{\mu\nu\sigma}_{(3)}$ and
$\gamma^\mu \gamma^\nu \gamma^\sigma$ and so on. Moreover, the iteration is 
trivial to implement symbolically and the output converted into the 
appropriate substitution routine. Once in this $\Gamma_{(n)}$-basis any 
spinor trace which is left can be carried out by noting that 
\begin{equation} 
\mbox{tr} \, \Gamma^{(n)} ~=~ 0 ~~~ \mbox{for $n$ $\neq$ $0$} 
\end{equation} 
and $\mbox{tr} \, \Gamma^{(0)}$ $=$ $2$. In practice, though, the performance 
is more efficient if we take traces at an earlier point.  

Although this completes the algorithm for performing the $\gamma$-algebra of
the basic three loop graphs, the generation of the evanescent operators means
that when they are substituted into Feynman diagrams the presence of 
$\Gamma_{(n)}$ for $n$ $\neq$ $1$ needs to be handled. First, one could
decompose each $\Gamma_{(n)}$ via (\ref{gamndef}), and then use the above 
algorithm to reconvert into the $\Gamma_{(n)}$ basis after loop integration.
However, this is not efficient, especially when one is dealing with the 
decomposition of $\Gamma_{(5)}$ since it will involve $120$ terms which with 
its tensor partner is beginning to generate a sizeable number of terms before 
one considers the number of Feynman diagrams for a topology or indeed the 
number of vertices in each. The more efficient way to proceed is to exploit 
another $\Gamma_{(n)}$ identity which contains the first equation of
(\ref{gamprod}) as a simple case. It is, \cite{15,27}, 
\begin{equation} 
\Gamma^{\mu_1 \ldots \mu_n}_{(n)} \Gamma^{\nu_1 \ldots \nu_n}_{(m)} 
\Gamma_{\mu_1 \ldots \mu_n}^{(n)} ~=~ f(n,m) \Gamma^{\nu_1 \ldots \nu_n}_{(m)} 
\label{gamnprod} 
\end{equation} 
where\footnote{We note that this relation appears in both \cite{27} and 
\cite{21}. In \cite{21} there appears to be a minor typographical error in a 
phase factor in the first equations of (A.1.2) and (A.1.3). The correct factor,
which we have used, ought to be $\varepsilon_m$ and not $\varepsilon_k$ which 
can be readily checked by evaluating special cases of $m$ and $n$.}
\begin{equation} 
f(n,m) ~=~ (-1)^{nm} (-1)^{n(n-1)/2} \, \left. \frac{\partial^n}{\partial u^n} 
\left[ (1+u)^{d-m} (1-u)^m \right] \right|_{u = 0} ~. 
\label{fnmdef} 
\end{equation}  
At the outset any diagram involving the inclusion of an evanescent operator
will be a mix of $\Gamma_{(n)}$ and ordinary $\gamma$'s. The latter, however,
are decomposed into $\Gamma_{(n)}$'s by the earlier algorithm before 
(\ref{gamnprod}) is applied at the end. Having ensured that the 
$\gamma$-algebra can be handled in terms of the basis $\Gamma_{(n)}$ we
note that the most general set of $4$-fermi operators covering the NATM is  
\begin{equation}  
\frac{1}{2} \left( \bar{\psi}^{iI} \delta_{IJ} \Gamma_{(n)} \psi^{iJ} 
\right)^2 ~~~,~~~ \frac{1}{2} \left( \bar{\psi}^{iI} \Gamma_{(n)} T^a_{IJ} 
\psi^{iJ} \right)^2 ~.  
\end{equation}  
One recovers this basis in practical terms by inverting (\ref{cvit}) in 
conjunction with the spinor indices of the $\Gamma_{(n)}$'s. For instance, if 
one is dealing with the four point function $\la \psi^{iI}_\alpha 
\bar{\psi}^{jJ \, \beta} \psi^{kK}_\gamma \bar{\psi}^{lL \, \delta} \ra$, then
\begin{equation}  
\delta_{IL} \delta_{KJ} \Gamma^{(n) ~ \beta}_{~~~ \alpha} \,  
\Gamma_{(n) \, \gamma}^{~~~~\, \delta} ~=~ \left[ 2 T^a_{IJ} T^a_{LK} ~+~ 
\frac{\delta_{IJ} \delta_{KL}}{\Nc} \right] \Gamma^{(n) ~ \beta}_{~~~ \alpha} 
\, \Gamma_{(n) \, \gamma}^{~~~~\, \delta} ~.  
\label{tdecomp} 
\end{equation}  

The other major part of the calculation is the integration of the Feynman 
diagrams which is based on the earlier papers of \cite{17,18}, which we follow 
here. We will compute each of the $2$-point and $4$-point functions in both the 
Gross Neveu model and NATM (massive and massless). The former will be at 
non-zero external momentum whilst for the $4$-point function we will set all 
external momenta to zero. This is another reason for the presence of a mass 
term which is necessary to have a scale in the integrals. In \cite{18} the 
diagrams were computed in $d$-dimensions with the substitution $d$ $=$ $2$ $-$ 
$\epsilon$ only being made at the end. We follow this source here primarily so 
that we can check that each of our integral programmes gives the correct result
in the Gross Neveu model before applying to the NATM. There is one complication
though. In the Gross Neveu model the vertex involves the spinor identity. 
Therefore within a topology the $\gamma$-matrices contracted with the momenta 
are adjacent. This simplified the Gross Neveu calculation as one did not need 
to consider tensor Feynman integrals. These, however, are necessary in the NATM
and we have had to rework the results of \cite{18} to determine the tensorial 
nature of the integrals which arise for each topology. By way of illustration 
we recall that a typical integral is 
\begin{equation} 
i^3 \int_{klq} \frac{k^\mu k^\nu k^\sigma l^\rho q^\lambda q^\psi}{(k^2-m^2)^2
((k-l)^2-m^2)(l^2-m^2)(q^2-m^2)^2} ~.  
\label{tenintdef} 
\end{equation} 
The most general form that can be consistent with the integral structure is 
\begin{equation} 
\eta^{\lambda\psi} \left[ X \eta^{\mu\nu} \eta^{\sigma\rho} ~+~  
Y \eta^{\mu\sigma} \eta^{\nu\rho} ~+~ Z \eta^{\mu\rho} \eta^{\nu\sigma} 
\right] ~.  
\end{equation} 
However, by Lorentz symmetry it is simple to deduce that $X$ $=$ $Y$ $=$ $Z$.
Hence the unknown amplitude is readily deduced by contraction with the 
resulting integral determined from the results in the appendix of \cite{18}. 
It evaluates to  
\begin{equation} 
\eta^{\lambda\psi} \left[ \eta^{\mu\nu} \eta^{\sigma\rho} ~+~  
\eta^{\mu\sigma} \eta^{\nu\rho} ~+~ \eta^{\mu\rho} \eta^{\nu\sigma} \right] 
\left[ I^2 ~+~ \frac{(d+3)}{3} m^2 \Delta(0) \right] \frac{I}{4d(d+2)} ~.  
\label{tenint} 
\end{equation} 
Throughout the computation two basic integrals emerge which are present in
(\ref{tenint}). The first is the simple tadpole result  
\begin{equation} 
I ~=~ i \int_k \frac{1}{(k^2-m^2)} 
\end{equation} 
which is divergent in two dimensions. The second term $\Delta(0)$ is the zero
momentum value of the more general integral 
\begin{equation} 
\Delta(p) ~=~ i \int_k \frac{J(k^2)}{((k-p)^2 - m^2)} 
\end{equation} 
where 
\begin{equation} 
J(p) ~=~ i \int_k \frac{1}{(k^2-m^2)((k-p)^2 - m^2)} ~.  
\end{equation} 
We leave $\Delta(0)$ unevaluated since it is connected to the issue of 
renormalizability. If we were performing the calculation of the $4$-point
function at non-zero external momenta the function $\Delta(p)$ as well as other
finite momentum dependent functions would emerge. For the theory to be 
renormalizable such quantities must not appear in the full evaluation of the 
Green's function multiplied by poles in $\epsilon$. If they did then they would
correspond to some non-local counterterm in the Lagrangian which would not fit 
the usual criteria for renormalizability. It occurs in a subset of the two and 
three loop topologies. The appearance at two loops is as a finite term but it 
will contribute at three loops when multiplied by a one loop counterterm. In 
\cite{17,18} their cancellation to give a finite contribution in the full 
expression for the $2$ and $4$-point functions was verified to ensure that the 
Gross Neveu model is renormalizable to three loops. We have checked this again 
within the context of our semi-automatic calculation before verifying that in 
the NATM and ABTM the sum of all $\Delta(0)/\epsilon$ pieces yield a finite 
contribution as it ought to, to preserve the usual renormalizability criteria 
at three loops. Since we will be considering non-minimal renormalization 
schemes in a later section we will require knowledge of the finite parts in 
addition to the poles in $\epsilon$. Therefore we have been careful in the 
construction of all the tensor integrals similar to (\ref{tenint}) not to 
discard any $d$-dependence. In other words even though say $(d-2)^2I^2$ is 
clearly finite in two dimensions we have not omitted it from any amplitude in 
the tensor decomposition.  
\vspace{0.5cm} 
\begin{figure}[hb] 
\epsfig{file=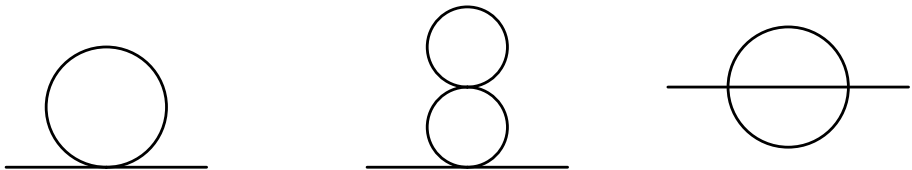,width=12cm} 
\vspace{0.5cm} 
\caption{One and two loop topologies for the $2$-point function.} 
\end{figure} 
\vspace{0.5cm} 
\begin{figure}[hb] 
\epsfig{file=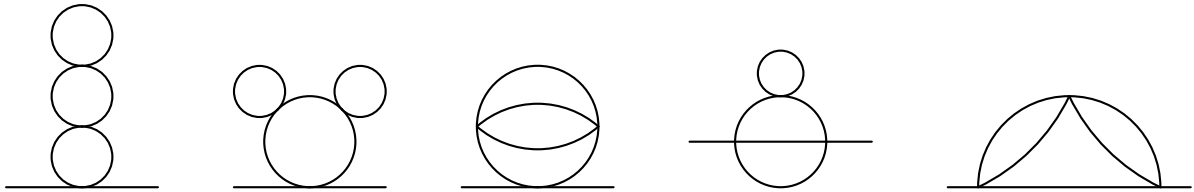,height=2.4cm} 
\vspace{0.5cm} 
\caption{Three loop topologies for the $2$-point function.} 
\end{figure} 

Finally, we recall that the topologies for the various diagrams can be readily
constructed. To three loops the basic ones for the $2$-point function are given
in figures 1 and 2. We have not labelled the graphs with arrows for the charge 
flow as to include all possibilites would increase the number of graphs 
significantly. Suffice to say that at one and two loops there are a total of 
$24$ graphs for the $2$-point function and at three loops a total of $7$ 
graphs. For the $4$-point function the topologies are illustrated in figures 3,
4 and 5. In these cases there are more graphs due to the existence of various 
scattering channels which are not illustrated. In total there are $196$ one and
two loop graphs and $213$ three loop graphs. For both Green's function these 
figures for one and two loops take into account the contributions from diagrams
with $2$-point counterterms and evanescent vertices. We have not counted as
separate those graphs generated by the vertex counterterm. 

\sect{Gross Neveu model.} 

Before considering the renormalization of the NATM in detail we address the 
multiplicative renormalizability issue of the Gross Neveu model. In 
constructing our algorithm we have been careful to check that the 
$d$-dimensional expression for each set of graphs contributing to the same 
topology agree with those given in \cite{18}. Although that calculation was 
carried out for the model with $O(\Nf)$ Majorana fermions, the observation that
$SU(\Nf)$ $=$ $O(2\Nf)$ allows us to compare the results. Moreover, the 
comparison is made only for the $\Gamma_{(0)} \otimes \Gamma_{(0)}$ type 
vertex. In \cite{18} and \cite{19} it was assumed that there were no other 
structures available which had not been justified. However, if new divergent 
structures did appear at three loops, they would not invalidate the three loop 
$\MSbar$ renormalization group functions computed in \cite{17,18,19}. With the 
machinery of the $\Gamma_{(n)}$-basis now available we have recomputed the full
$4$-point Green's function in the $SU(\Nf)$ Gross Neveu model. It turns out 
that at three loops several topologies give rise to non-zero contributions 
involving $\Gamma_{(3)} \otimes \Gamma_{(3)}$ both for the massive and massless
models. It could, of course, be the case that their sum is finite in two 
dimensions. However, we first record the values of the $\Gamma_{(3)} \otimes 
\Gamma_{(3)}$ part of the contributing graphs as  
\begin{eqnarray} 
G_{41}(\OO_3) &=& \frac{I^3g^4}{4} ~~~,~~~ 
G_{42}(\OO_3) ~=~ -~ \frac{I^3 g^4}{6d} ~~~,~~~  
G_{44}(\OO_3) ~=~ \frac{(3d-4) I^3 g^4}{12d(d-1)^2} \nonumber \\  
G_{45}(\OO_3) &=& -~ \frac{I^3 g^4}{2d} ~~~,~~~  
G_{46}(\OO_3) ~=~ -~ \frac{(3d-4) I^3 g^4}{6d(d-1)^2} ~~~,~~~  
G_{47}(\OO_3) ~=~ \frac{I^3 g^4}{3d(d-1)} 
\end{eqnarray}  
\begin{figure}[ht] 
\epsfig{file=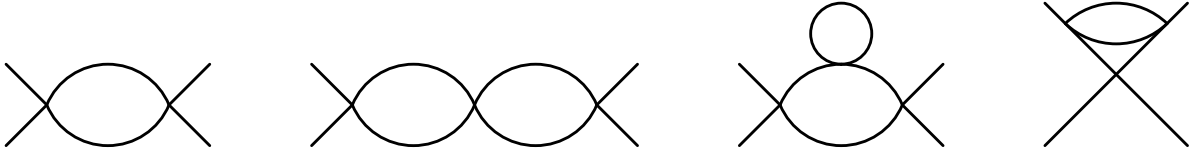,height=1.9cm} 
\vspace{0.5cm} 
\caption{One and two loop topologies for the $4$-point function.} 
\end{figure} 
where the subscripts correspond to the graphs of figure 4. It is worth 
observing that these extra contributions in the $\Gamma_{(3)} \otimes 
\Gamma_{(3)}$ sector only arise in those diagrams with {\em vertex} subgraphs. 
It is easy to sum the contributions of the graphs to find the total is  
\begin{equation} 
G^{(4)}(\OO_3) ~=~ \frac{(3d-2)(d-2)^2I^3g^4}{12d(d-1)^2} 
\end{equation} 
which possesses a simple pole in $\epsilon$. Therefore, we need to include an
extra set of evanescent operator counterterms in the original Lagrangian, 
(\ref{gnlag}), which involved only bare fields and parameters. So, if the bare
Lagrangian is 
\begin{equation} 
L^{\mbox{\footnotesize{gn}}} ~=~ i \bar{\psi}_0^{i} \partialslash \psi^{i}_0 
{}~-~ m_0 \bar{\psi}^{i}_0 \psi^{i}_0 ~+~ \half g_0 ( \bar{\psi}^i_0 
\psi^i_0 )^2
\end{equation}  
then we introduce the Lagrangian involving renormalized fields and parameters 
as 
\begin{equation} 
L^{\mbox{\footnotesize{gn}}} ~=~ i Z_\psi \bar{\psi}^{i} \partialslash \psi^{i}
{}~-~ m Z_\psi Z_m \bar{\psi}^{i} \psi^{i} ~+~ \half g \mu^\epsilon 
Z_g Z^2_\psi ( \bar{\psi}^i \psi^i )^2 ~+~ \half g \mu^\epsilon Z_{33} 
Z^2_\psi \left( \bar{\psi}^i \Gamma_{(3)} \psi^i \right)^2 
\label{gnevlag} 
\end{equation}  
which will be valid to three loops and where we define the usual 
renormalization constants by 
\begin{equation} 
\psi_0 ~=~ \psi Z_\psi^{\half} ~~~,~~~ m_0 ~=~ m Z_m ~~~,~~~ g_0 ~=~ 
g Z_g \mu^\epsilon ~.  
\end{equation} 
We have checked that $Z_\psi$, $Z_m$ and $Z_g$ are in agreement with the 
$\MSbar$ results of \cite{17,18,19}. Additionally, we have 
\begin{equation} 
Z_{33} ~=~ -~ \frac{g^3}{48\pi^3\epsilon} ~. 
\label{gnz33} 
\end{equation} 

The form of $Z_{33}$ is typical of the structure of renormalization constants
associated with evanescent operators in that its leading term is not $O(1)$. 
This can be appreciated by considering the Lagrangian analogous to 
(\ref{abtmevlag}) where one begins with an infinite set of basis evanescent 
operators each with their own coupling. If one were to renormalize with that 
theory then the resulting renormalization constants would have a similar 
structure to $Z_g$ with the corrections involving all the couplings. However, 
in the limit $g_k$ $\rightarrow$ $0$, it is simple to see that the form 
(\ref{gnz33}) would emerge. With (\ref{gnz33}) our observation is that the 
$SU(\Nf)$ Gross Neveu model is not multiplicatively renormalizable at three 
loops in dimensional regularization. Moreover, the lowest operators that can be
generated in principle is $\Gamma_{(3)} \otimes \Gamma_{(3)}$. This is due to 
the fact that if $\Gamma_{(1)} \otimes \Gamma_{(1)}$ or $\Gamma_{(2)} \otimes 
\Gamma_{(2)}$ appeared one would have to include counterterms for operators
which do not vanish in the limit $d$ $\rightarrow$ $2$. This would be counter
to the multiplicative renormalization in other strictly two dimensional 
regularizations. If we count the total possible number of $\gamma$-matrices 
which will occur in a $4$-point function at successive loops up to three we 
find that these will be respectively $2$, $4$ and $6$. It is clear therefore
that the first appearance of a truly evanescent operator cannot occur before 
three loops which is what we find. 
\vspace{0.5cm} 
\begin{figure}[ht]
\epsfig{file=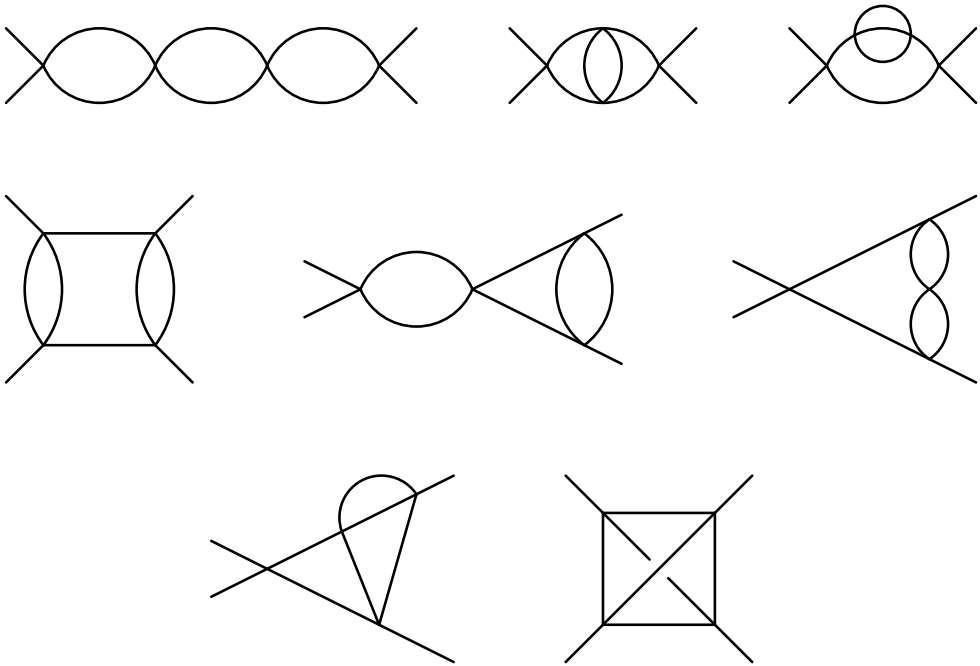,width=12.9cm} 
\vspace{0.5cm} 
\caption{Basic three loop topologies for the $4$-point function.} 
\end{figure} 

As will be discussed in more detail later for the NATM, the existence of such
an operator only affects the renormalization of the model at the order after
which it first appears. This is because the extra term in (\ref{gnevlag}) gives
rise to a new vertex with a coupling involving $Z_{33}$. Therefore the first 
diagrams where (\ref{gnz33}) will be relevant are given in figure 6 where we 
denote the location of the $4$-point evanescent operator by the cross in a 
circle. We note that one needs to take care when counting the $g$ dependence of
such graphs, since $Z_{33}$ is $O(g^3)$. Whilst the anomalous dimension 
$\gamma(g)$ has been computed to $4$-loops in \cite{38} in $\MSbar$ that result
does not require the $\Gamma_{(3)} \otimes \Gamma_{(3)}$ operator insertion in 
the $2$-point function. It is easy to see this since the first Feynman 
integral, when computed, only contributes to the mass dimension. The integral 
involving the loop momentum in the numerator vanishes by Lorentz symmetry. Thus
only the $4$-loop $\beta$-function and mass anomalous dimension will require 
the computation of the graphs of figure 6 in addition, of course, to the 
(large) set of diagrams arising from the original vertex. The detailed 
procedure to achieve this will be discussed in the context of the NATM where 
the same problem arises but at a lower loop order, \cite{15}.  
\vspace{0.5cm} 
\begin{figure}[ht] 
\epsfig{file=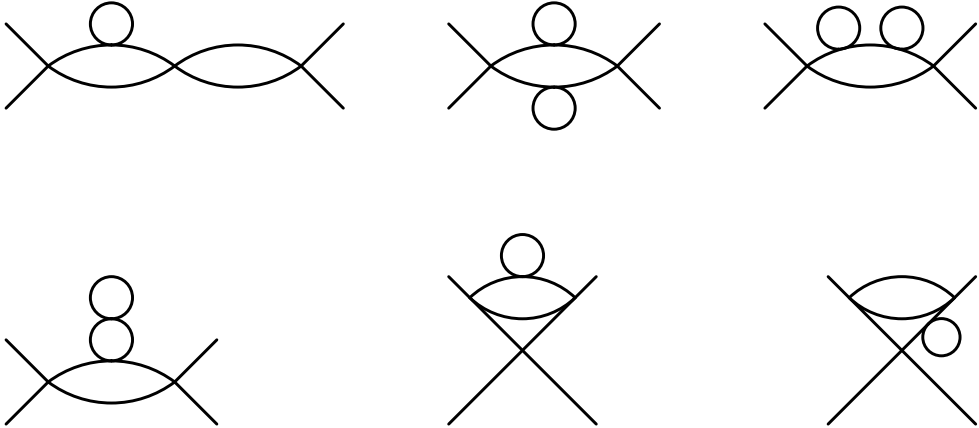,width=12.9cm} 
\vspace{0.5cm} 
\caption{Three loop topologies with tadpoles for the $4$-point function.} 
\end{figure} 
\vspace{0.5cm} 
\begin{figure}[hb]  
\epsfig{file=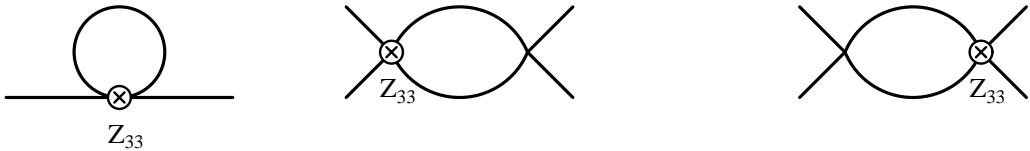,height=2cm} 
\vspace{0.5cm} 
\caption{Contributions from evanescent operators to the four loop wave 
function, mass and coupling constant renormalization in the Gross Neveu model.} 
\end{figure} 

\sect{Massive NATM in $\MSbar$.} 

As far as we aware there been two calculations dealing with the two loop 
renormalization of the (massless) NATM, \cite{14,15}. The initial one performed
in \cite{14} used the auxiliary field formulation with a cutoff regularization 
which is not easy to extend to three loops. The latter one was detailed in the 
appendix of \cite{15} and used dimensional regularization. Using the elementary
argument of the previous section to ascertain the form of the highest possible 
operator which could emerge from the $4$-point function, it is elementary to 
observe that a $\Gamma_{(3)} \otimes \Gamma_{(3)}$ structure can emerge at one 
loop. This is evident in \cite{15} and as a preliminary to discussing the three
loop calculation we will first reproduce the analysis of \cite{15} in the 
context of this article. There will be several minor differences in our 
strategy. For instance, we will be interested in the mass renormalization which
was not discussed before. Second, the results presented in \cite{15} were for a 
general colour group and did not appear to appeal to the simplification 
manifested by (\ref{tdecomp}) for $SU(\Nc)$. Although this is not necessary
to two loops the problem of obtaining independent combinations of the colour 
generators is not trivial in the general case, \cite{35}. Indeed it is tedious 
even at the two loop level. At the outset we note that any calculation we have 
performed at two loops has been checked with the NATM analysis of \cite{15} 
including the coefficients of the renormalization constants of the usual 
parameters and emerging evanescent operators, as well as the simple application
of the projection formula to obtain the correct $\beta$-function of \cite{14}.
Although we will extend the massless calculation of \cite{15} to three loops, 
our strategy will be to first consider the massive model. The primary reason
for doing this is to exploit non-trivial checks on the full renormalization
group functions which become available in the massive model. Having understood
the procedure for the massive model we proceed to the massless case where
such a check does not exist. This is due to the different roles $m$ plays in
both models. In the latter case, since $m$ is a regularizing parameter it does
not appear in the full renormalization group equation.  

We begin by defining the renormalization constants we will compute, where the
bare Lagrangian is 
\begin{equation} 
L^{\mbox{\footnotesize{natm}}} ~=~ i \bar{\psi}_0^{iI} \partialslash 
\psi_0^{iI} ~-~ m_0 \bar{\psi}_0^{iI} \psi_0^{iI} ~+~ \frac{g_0}{2}   
\left( \bar{\psi}_0^{iI} \gamma^\mu T^a_{IJ} \psi_0^{iJ} \right)^2 ~. 
\end{equation}  
Setting 
\begin{equation} 
\psi_0 ~=~ \psi Z_\psi^{\half} ~~~,~~~ m_0 ~=~ m Z_m ~~~,~~~ g_0 ~=~ g Z_{11}  
\mu^\epsilon 
\end{equation} 
and allowing for all possible evanescent operators up to three loops, we will
take the renormalized Lagrangian as 
\begin{eqnarray} 
L^{\mbox{\footnotesize{natm}}} &=& i Z_\psi \bar{\psi}^{iI} \partialslash 
\psi^{iI} ~-~ m Z_m Z_\psi \bar{\psi}^{iI} \psi^{iI} \nonumber \\ 
&& +~ \frac{1}{2} g \mu^\epsilon Z^2_\psi \sum_{k = 0}^7 \left[ Z_{k0} \left( 
\bar{\psi}^{iI} \Gamma_{(k)} \delta_{IJ} \psi^{iJ} \right)^2 ~+~ Z_{k1} \left( 
\bar{\psi}^{iI} \Gamma_{(k)} T^a_{IJ} \psi^{iJ} \right)^2 \right]  
\end{eqnarray}  
where as in (\ref{gnevlag}) the $Z_{ki}$ are $O(g)$ except when $k$ $=$ $1$ and
$i$ $=$ $1$. Although it would appear that there are thirteen terms up to level
$\Gamma_{(7)} \otimes \Gamma_{(7)}$ in addition to the original vertex, it 
turns out that only the operators involving $\Gamma_{(2r+1)}$ for integer $r$,
will occur, \cite{21,22}, and the explicit calculation is consistent with this.
To one loop the relevant diagrams are given in figures 1 and 3. The tadpole, 
defined as a zero momentum insertion on a line, only contributes to the mass 
renormalization, and gives the value  
\begin{equation} 
\frac{i(\Nc^2-1)dmIg}{2\Nc} ~.  
\label{two1} 
\end{equation} 
For the $4$-point function the sum of all channels gives  
\begin{eqnarray} 
&& \frac{i g^2 I}{4} \left[ ~-~ \frac{(\Nc^2-1)}{\Nc^2} d(d-2) \OO_{00} ~-~ 
\frac{(\Nc^2-4)}{\Nc} d(d-2) \OO_{01} ~-~ \Nc (3d-2) \OO_{11} \right. 
\nonumber \\ 
&& \left. ~~~~~~~~~+~ \Nc (d-2) \OO_{21} ~+~ \frac{(\Nc^2-1)}{\Nc^2} 
\OO_{30} ~+~ \frac{(\Nc^2-4)}{\Nc} \OO_{31} \right] ~.  
\label{four1} 
\end{eqnarray}  
Here, we adopt the practice that the Green's functions are written in terms of 
the (basis) operators $\OO_{n0}$ $\equiv$ $\half (\bar{\psi} \Gamma_{(n)} I 
\psi)^2$ and $\OO_{n1}$ $\equiv$ $\half (\bar{\psi} \Gamma_{(n)} T^a \psi)^2$ 
whose definition does not contain any coupling constant dependence. By doing 
this one can easily define the counterterm appropriately. At this stage we 
choose to renormalize using the $\MSbar$ scheme and absorb only those parts of
(\ref{two1}) and (\ref{four1}) which are divergent. We have computed the
coefficients of each operator as a function of $d$ in order to consider other 
schemes which we will do later. Therefore due to the presence of the factors of
$(d-2)$ in (\ref{four1}) counterterms are only required for $\OO_{11}$, 
$\OO_{30}$ and $\OO_{31}$ which have an odd number of $\gamma$-matrices. This 
is consistent with the earlier observation relating to chirality, \cite{21,22}.
The generation of the latter operators affects the two loop renormalization. By 
counting the powers of the coupling constant in the Lagrangian, they need to be
included with the two loop diagrams of figures 1 and 3, unlike in the Gross 
Neveu model where they contribute at four loops. Therefore, in the $2$-point 
function they are only relevant for the mass dimension through the tadpole of 
figure 6. For the $4$-point function of figure 6, the insertion now represents 
the presence of both $\OO_{30}$ and $\OO_{31}$. 

At two loops in the $2$-point function there will be a contribution to the wave
function renormalization from the sunset diagram. Of all the graphs we
consider, this, aside from its related graphs at three loops, represented the
most tedious to evaluate. It was not possible to write the relevant integral in
a closed form. Instead we treated it by expanding the integral in powers of 
$p^2$ around $p^2$ $=$ $0$ and retained only those pieces which involve 
$\pslash$ or $m\delta_\alpha^{~\,\beta}$. These contributed respectively to the
wave function and mass renormalization. In addition, we were careful to 
determine not only the pole parts but also the finite parts for the three loop
calculation and in order to consider non-minimal renormalization schemes. We 
note that the wave function and other renormalization constants to three loops 
in the $\MSbar$ scheme have been collected together in appendix A. 

Computing the $4$-point function now yields in addition to the operators
$\OO_{30}$ and $\OO_{31}$, the new operators $\OO_{10}$, $\OO_{50}$ and 
$\OO_{51}$. The latter two would be expected due to the simple $\gamma$-matrix
counting, since the addition of a vertex introduces two $\gamma$-matrices from 
the new propagators and two $\gamma$-matrices from the extra vertex. For the 
one loop diagram of figure 6, the presence of $\Gamma_{(3)}$ at the operator 
insertion also yields a maximum of ten $\gamma$-matrices which have the extreme
decomposition of $\Gamma_{(5)} \otimes \Gamma_{(5)}$. The main issue with the
two loop $4$-point function is the observation that one cannot simply deduce 
the full $\beta$-function from the na\"{\i}ve renormalization constant 
$Z_{11}$. Ordinarily in the absence of the higher order operators the 
coefficients of the simple poles in $\epsilon$ are simply related to the 
coefficients of the $\beta$-function. In this case we would have  
\begin{equation} 
\tilde{\beta}(g) ~\equiv~ \tilde{\beta}_{11}(g) ~=~ \frac{\Nc g^2}{2\pi} ~+~ 
(4\Nc^3\Nf - 3\Nc^4 + 12\Nc^2 - 36) \frac{g^3}{16\Nc^2\pi^2} ~+~ O(g^4) ~.  
\end{equation} 
As was pointed out in \cite{15}, this clearly contradicts the calculation of
\cite{14} since on general considerations the $\beta$-function for a single 
coupling theory is scheme independent to two loops. The discrepancy between the
two results is $3(\Nc^4-4\Nc^2+12)g^3/(16\Nc^2)$ which is related to the 
$\beta$-function of the operator ${\cal F}$ in the notation of \cite{15} where 
\begin{equation} 
{\cal F} ~=~\half \bar{\psi} T^a T^b T^c \gamma^\mu \psi \bar{\psi} 
{\cal T}^{abc} \gamma_\mu \psi 
\end{equation} 
with 
\begin{equation} 
{\cal T}^{abc} ~=~ 2 T^a T^b T^c ~+~ 2 T^c T^b T^a ~+~ T^a T^c T^b ~+~ 
T^b T^a T^c ~+~ T^c T^a T^b ~+~ T^b T^c T^a 
\end{equation} 
and 
\begin{equation}  
\beta_{{\cal F}}(g) ~=~ -~ \frac{3g^3}{16\pi^2} ~.  
\end{equation}  
For $SU(\Nc)$, 
\begin{equation} 
\left. {\cal F} \right|_{SU(\Nc)} ~=~ \frac{1}{2} \left[ 
\frac{(\Nc^2-1)(\Nc^2-4)}{2\Nc^3} ( \bar{\psi} \gamma^\mu I \psi )^2 ~+~  
\frac{(\Nc^4-4\Nc^2+12)}{2\Nc^2} ( \bar{\psi} \gamma^\mu T^a \psi )^2 \right] 
\end{equation} 
so that the second term is related to the spurious contribution to the 
na\"{\i}ve $\beta$-function. Clearly one needs a systematic method of 
accounting for the existence of such contributions and removing them from the 
true $\beta$-function and the other renormalization group functions. 

This is provided for in the projection technique, \cite{28}, discussed in 
\cite{15} but which we recast in the basis of operators we have chosen. In 
addition to the na\"{\i}ve $\beta$-function, $\tilde{\beta}_{11}(g)$, the 
evanescent operators generated at one loop also have $\beta$-functions, deduced
in the usual fashion. To $O(g^3)$ these are 
\begin{eqnarray} 
\tilde{\beta}_{30}(g) &=& -~ \frac{(\Nc^2-1)g^2}{8\Nc^2\pi} \\  
\tilde{\beta}_{31}(g) &=& -~ \frac{(\Nc^2-4)g^2}{8\Nc\pi} ~.  
\end{eqnarray} 
Although the general formalism was presented in section 2 its practical
application requires explanation, especially as it will be applied to the three
loop renormalization. As it stands, equation (\ref{projdef}) represents the 
decomposition of the evanescent operator into a linear combination of the
relevant original operators. The coefficients of this combination are 
perturbative functions of the physical coupling and the formula is only 
meaningful in the context of a Green's function. Substituting, say, the 
operator $\NN[\OO_{31}]$ into a $2$-point function one evaluates the Green's 
function in two dimensions {\em after} any infinities have been removed. Then 
one inserts the right side of (\ref{projdef}) into the same Green's function 
with the as yet undetermined projection functions and evaluates the Feynman 
diagrams to the same order before renormalizing and setting $d$ $=$ $2$. 
However, it is easy to deduce that at leading order only the operator 
$\NN[\OO_{11}]$ will contribute for insertions in a $4$-point function. 
Likewise $\NN[i\bar{\psi} \partialslash \psi]$ and $\NN[\bar{\psi} \psi]$ will 
be relevant for the leading order in the $2$-point function. This therefore 
provides the starting point of the perturbative iteration which determines 
$\rho^{(ki)}(g)$, $\rho^{(ki)}_m(g)$ and $C^{(ki)}(g)$, $i$ $=$ $0$, $1$. 
Indeed at leading order it is only the tree level insertions of the relevant 
operators on the right side of (\ref{projdef}) which will contribute. The 
insertion of these in the respective Green's function each evaluate to unity. 

Hence one needs only to compute the graphs of figure 6 for each operator 
$\OO_{30}$ and $\OO_{31}$. For the $2$-point function each is finite to this
order and give  
\begin{eqnarray} 
\la \psi \NN [ \OO_{30} ] \bar{\psi} \ra_{wf} &=& O(g) \nonumber \\  
\la \psi \NN [ \OO_{31} ] \bar{\psi} \ra_{wf} &=& O(g) 
\end{eqnarray} 
and 
\begin{eqnarray} 
\la \psi \NN [ \OO_{30} ] \bar{\psi} \ra_m &=& \frac{1}{\pi} ~+~ O(g) 
\nonumber \\  
\la \psi \NN [ \OO_{31} ] \bar{\psi} \ra_m &=& \frac{(\Nc^2-1)}{2\Nc\pi} ~+~ 
O(g) ~.  
\end{eqnarray} 
We have separated the contributions to the full $2$-point function with the
operator insertion into the piece involving $\pslash$ and the mass and denoted
these respectively by the subscripts $wf$ and $m$. However, for the $4$-point 
function care must be taken to renormalize the operator insertion before 
setting $d$ $=$ $2$. As this will become important later we record the 
insertion of each operator in the $4$-point function to the $O(\epsilon)$ term 
\begin{eqnarray} 
\la \psi \bar{\psi} \NN [ \OO_{30} ] \psi \bar{\psi} \ra &=& \frac{ig}{\pi} 
\left[ \frac{1}{\epsilon} \left( \OO_{51} ~-~ \frac{4(\Nc^2-1)}{\Nc} \OO_{30} 
\right) ~-~ 6 \OO_{11} \right. \nonumber \\ 
&& \left. ~~~-~ \frac{(\Nc^2-1)}{\Nc} \left( 1 + 2 \ln \left( 
\frac{\tilde{\mu}^2}{m^2} \right) \right) \OO_{30} ~+~ \frac{1}{2} \ln \left( 
\frac{\tilde{\mu}^2}{m^2} \right) \OO_{51} \right. \nonumber \\ 
&& \left. ~~~+~ 3 \epsilon \left( 2 - \ln \left( \frac{\tilde{\mu}^2}{m^2} 
\right) \right) \OO_{11} ~+~ O(\epsilon;\OO_{k0},\OO_{k1}) \right] ~+~ O(g^2)  
\label{o30} 
\end{eqnarray} 
and 
\begin{eqnarray} 
\la \psi \bar{\psi} \NN [ \OO_{31} ] \psi \bar{\psi} \ra &=& \frac{ig}{\pi} 
\left[ \frac{1}{\epsilon} \left( \frac{(\Nc^2+4)}{\Nc} \OO_{31} ~+~ 
\frac{(\Nc^2-1)}{4\Nc^2} \OO_{50} ~+~ \frac{(\Nc^2-4)}{4\Nc} \OO_{51} \right)  
\right. \nonumber \\ 
&& \left. ~~~-~ \frac{3(\Nc^2-1)}{2\Nc^2} \OO_{10} ~-~ \frac{3(\Nc^2-4)}{2\Nc} 
\OO_{11} ~-~ \frac{\Nc}{4} \OO_{41} \right. \nonumber \\ 
&& \left. ~~~+~ \left( 7\Nc^2 + 4 + 2(\Nc^2 + 4) \ln \left( 
\frac{\tilde{\mu}^2}{m^2} \right) \right) \frac{{\OO_{31}}}{4\Nc} \right. 
\nonumber \\ 
&& \left. ~~~+~ \frac{(\Nc^2-1)}{8\Nc^2} \ln \left( \frac{\tilde{\mu}^2}{m^2} 
\right) \OO_{50} ~+~ \frac{(\Nc^2-4)}{8\Nc} \ln \left( 
\frac{\tilde{\mu}^2}{m^2} \right) \OO_{51} \right. \nonumber \\ 
&& \left. ~~~+~ \frac{3(\Nc^2-4)\epsilon}{4\Nc} \left( 2 - \ln \left( 
\frac{\tilde{\mu}^2}{m^2} \right) \right) \OO_{11} ~+~ 
O(\epsilon;\OO_{k0},\OO_{k1}) \right] ~+~ O(g^2) \nonumber \\  
\label{o31} 
\end{eqnarray} 
where we have introduced the new mass scale $\tilde{\mu}^2$ $=$ 
$4\pi e^{-\gamma}\mu^2$, with $\gamma$ denoting Euler's constant, to ensure 
that we are in the $\MSbar$ scheme as opposed to the MS scheme. The 
renormalization constants to remove the simple poles on the right side are 
provided by the fact that the expressions effectively represent the first stage
in the renormalization of a composite operator in the normal fashion. There is 
a caveat here in that there is mixing into other basis operators. However, one 
can introduce a counterterm which is a linear combination of the basis 
elements. The upshot of this for higher order calculations is that when the 
renormalized operator is inserted one must also include the counterterms from 
the mixing operators. For instance, for $\OO_{30}$ this will mean at the next 
order that a tadpole graph with insertions $4(\Nc^2-1)\OO_{31}/(\Nc\epsilon)$ 
and $-$~$\OO_{51}/\epsilon$ have to be included. After this renormalization 
which we also take to be $\MSbar$ here, the contribution to the Green's 
function after setting the evanescent couplings to zero and $d$ $=$ $2$ are, to
this order,  
\begin{eqnarray} 
\left. \la \psi \bar{\psi} \NN [ \OO_{30} ] \psi \bar{\psi} \ra 
\right|_{\OO_{11}} &=& -~ \frac{6ig}{\pi} ~+~ O(g^2) \nonumber \\  
\left. \la \psi \bar{\psi} \NN [ \OO_{31} ] \psi \bar{\psi} \ra 
\right|_{\OO_{11}} &=& -~ \frac{3i(\Nc^2-4)g}{2\Nc\pi} ~+~ O(g^2) ~.  
\end{eqnarray} 
So it is trivial to deduce, 
\begin{eqnarray} 
\rho^{(30)}(g) &=& O(g) ~~~,~~~ \rho^{(31)}(g) ~=~ O(g) \\  
\rho^{(30)}_m(g) &=& -~ \frac{1}{\pi} ~+~ O(g) ~~~,~~~ \rho^{(31)}_m(g) ~=~ 
-~ \frac{(\Nc^2-1)}{2\Nc\pi} ~+~ O(g) \\  
C^{(30)}(g) &=& -~ \frac{6g}{\pi} ~+~ O(g^2) ~~~,~~~ 
C^{(31)}(g) ~=~ -~ \frac{3(\Nc^2-4)g}{2\Nc\pi} ~+~ O(g^2) ~.  
\end{eqnarray} 
Hence, substituting these functions into the general formul{\ae} in 
(\ref{truerge}) the actual renormalization group functions to two loops are 
\begin{eqnarray} 
\beta(g) &=& \frac{\Nc g^2}{2\pi} ~+~ \frac{\Nf\Nc g^3}{4\pi^2} \nonumber \\
\gamma(g) &=& -~ \frac{(\Nc^2-1)\Nf g^2}{8\Nc\pi^2} \nonumber \\
\gamma_m(g) &=& \frac{(\Nc^2-1) g}{2\Nc\pi} ~+~ (\Nc^2-1)(4\Nf - \Nc) 
\frac{g^2}{16\Nc \pi^2} 
\label{reg2loop} 
\end{eqnarray}  
where the result of \cite{14} is recovered. Moreover, we also have the $\MSbar$
mass anomalous dimension which has not been computed in this model before. 

We now turn to the details of the three loop calculations which builds on the
above. We do not comment on the mundane task of the computation of the 
diagrams of figures 2, 4 and 5 since their evaluation very much parallels any
usual perturbative calculation in other models. Instead we focus on the role
of the evanescent operators. As the two loop calculation generated new
operators their contributions must also be included. However, they will only 
enter through the topologies of figure 6 where the circle with a cross will
now represent successively $\OO_{10}$, $\OO_{50}$ and $\OO_{51}$. Additionally
the counterterms generated for $\OO_{30}$ and $\OO_{31}$ are also included in
those topologies as well as the original operator in the graphs of figures 7, 8
and 9. In figure 8 the insertion stands for the four possible combinations of 
$\OO_{30}$ and $\OO_{31}$. In these and the graphs 
%\vspace{0.5cm} 
\begin{figure}[ht]  
\epsfig{file=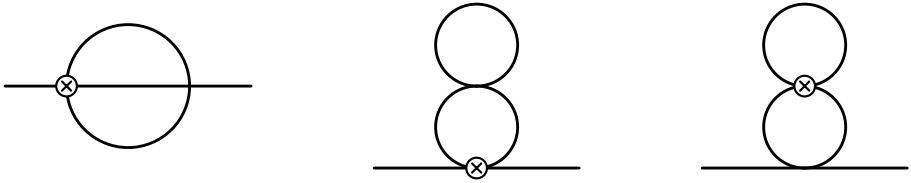,width=12cm} 
\vspace{0.5cm} 
\caption{Graphs contributing to the NATM $2$-point function from evanescent 
operators $\OO_{3i}$ in the two loop topology.} 
\end{figure} 
with the original vertex we note that we have also included the usual wave 
function, mass and vertex counterterms. Having detailed the extra evanescent 
operator contributions which have to be included it perhaps can be appreciated 
that such a calculation can only properly be attacked using symbolic and 
algebraic manipulation programmes.   
\vspace{0.5cm} 
\begin{figure}[hb] 
\epsfig{file=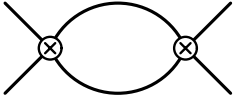,width=3cm} 
\vspace{0.5cm} 
\caption{Graphs contributing to the NATM $4$-point function from evanescent 
operators $\OO_{3i}$ in the one loop topology.} 
\end{figure} 

It is not possible to include the values of each diagram in terms of the full
operator basis. To illustrate the complexity of the result for one diagram,
though, we record the structure of the three loop chain of figure 4. We 
found that for arbitrary $d$ it is 
\begin{eqnarray} 
&& i g^4 I^3 \left[ ~-~ d(3d^3 + 31d^2 - 70d + 40)(d-2)(\Nc^4 - 2\Nc^2 + 6)
(\Nc^2 - 1) \frac{\OO_{00}}{128\Nc^4} \right. \nonumber \\ 
&& \left. ~~~~~~~~~-~ d(3d^3 + 31d^2 - 70d + 40)(d-2)(\Nc^4 + 2)(\Nc^2 - 4) 
\frac{\OO_{01}}{64\Nc^3} \right. \nonumber \\ 
&& \left. ~~~~~~~~~-~ (45d^4 - 165d^3 + 168d^2 + 36d - 80)(\Nc^2 - 4) 
(\Nc^2 - 1) \frac{\OO_{10}}{128\Nc^2} \right. \nonumber \\ 
&& \left. ~~~~~~~~~-~ (45d^4 - 165d^3 + 168d^2 + 36d - 80) 
(\Nc^4 - 4\Nc^2 + 10) \frac{\OO_{11}}{64\Nc} \right. \nonumber \\ 
&& \left. ~~~~~~~~~+~ (6d^3 + 103d^2 - 394d + 376)(d-2)(\Nc^2 - 4)(\Nc^2 - 1) 
\frac{\OO_{20}}{128\Nc^2} \right. \nonumber \\ 
&& \left. ~~~~~~~~~+~ (6d^3 + 103d^2 - 394d + 376)(d-2)(\Nc^4 - 4\Nc^2 + 10) 
\frac{\OO_{21}}{64\Nc} \right. \nonumber \\ 
&& \left. ~~~~~~~~~+~ (30d^3 - 75d^2 - 130d + 376)(\Nc^4 - 2\Nc^2 + 6) 
(\Nc^2 - 1)  \frac{\OO_{30}}{128\Nc^4} \right. \nonumber \\ 
&& \left. ~~~~~~~~~+~ (30d^3 - 75d^2 - 130d + 376)(\Nc^4 + 2)(\Nc^2 - 4)  
\frac{\OO_{31}}{64\Nc^3} \right. \nonumber \\ 
&& \left. ~~~~~~~~~-~ (d^2 + 41d - 116)(d-2)(\Nc^4 - 2\Nc^2 + 6)(\Nc^2 - 1) 
\frac{\OO_{40}}{128\Nc^4} \right. \nonumber \\ 
&& \left. ~~~~~~~~~-~ (d^2 + 41d - 116)(d-2)(\Nc^4 + 2)(\Nc^2 - 4) 
\frac{\OO_{41}}{64\Nc^3} \right. \nonumber \\ 
&& \left. ~~~~~~~~~-~ (3d^2 + 9d - 58) \left( (\Nc^2 - 4)(\Nc^2 - 1)  
\frac{\OO_{50}}{128\Nc^2} ~+~ (\Nc^4 - 4\Nc^2 + 10) \frac{\OO_{51}}{64\Nc} 
\right) \right. \nonumber \\ 
&& \left. ~~~~~~~~~+~ 3(d - 2) \left( (\Nc^2 - 4)(\Nc^2 - 1) 
\frac{\OO_{60}}{128\Nc^2} ~+~ (\Nc^4-4\Nc^2+10) \frac{\OO_{61}}{64\Nc} \right) 
\right. \nonumber \\ 
&& \left. ~~~~~~~~~+~ (\Nc^4 - 2\Nc^2 + 6)(\Nc^2 - 1)  
\frac{\OO_{70}}{128\Nc^4} ~+~ (\Nc^4 + 2)(\Nc^2 - 4) \frac{\OO_{71}}{64\Nc^3} 
\right] ~.  
\end{eqnarray}  
Unlike the one loop result for the graph of figure 3 the evanescent operators 
which are even in the number of $\gamma$-matrices have, in principle, been 
generated with simple and double poles in $\epsilon$. However, it turns out 
that when all contributions are summed and evaluated that they have finite 
coefficients. Thus divergences are only associated with those operators which 
have an odd number of $\gamma$-matrices as expected from general arguments, 
\cite{21,22}. Further, the new operators $\OO_{70}$ and $\OO_{71}$ appear which
is consistent with our earlier argument. Renormalizing the contributions 
\begin{figure}[ht] 
\epsfig{file=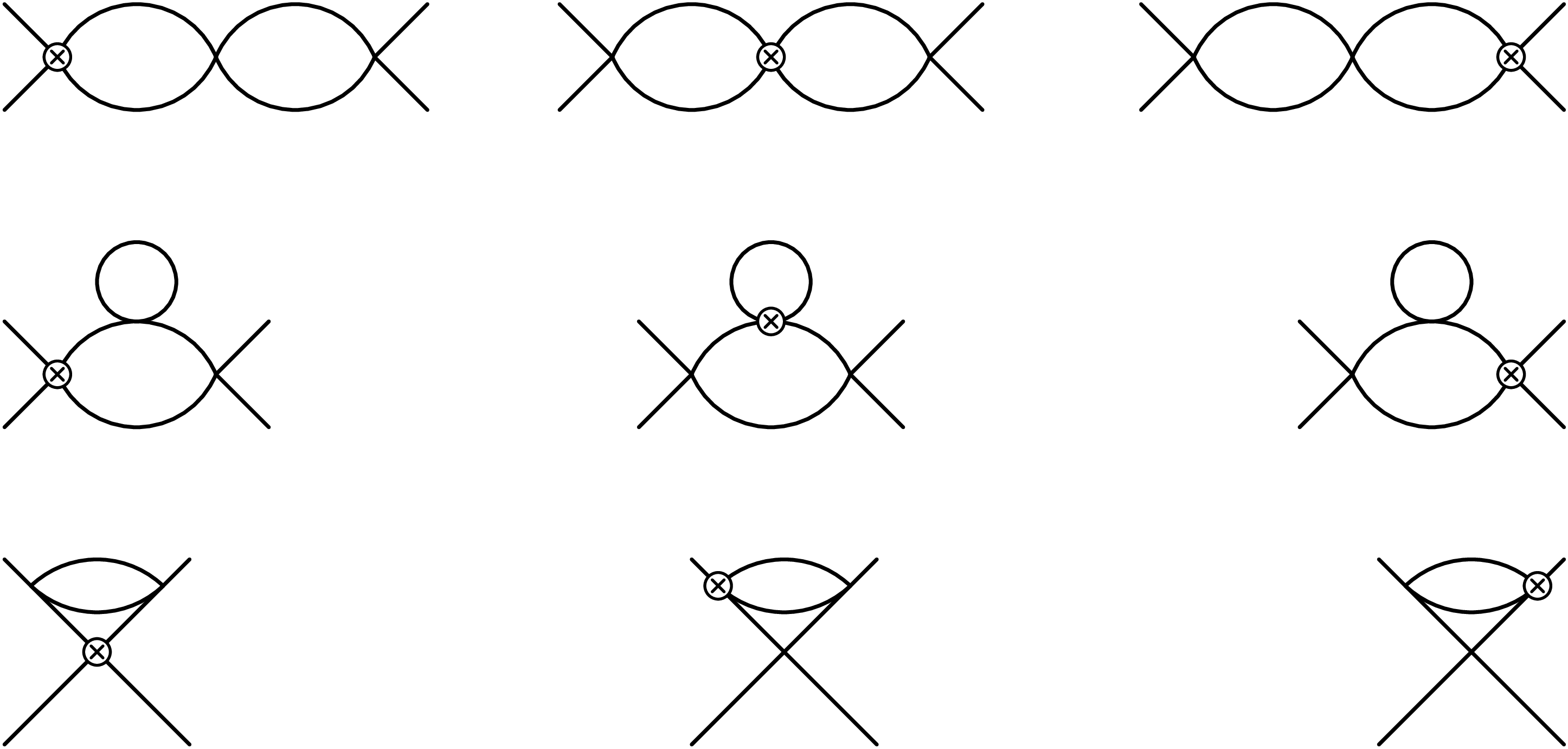,height=7.5cm} 
\vspace{0.5cm} 
\caption{Graphs contributing to the NATM $4$-point function from evanescent 
operators $\OO_{3i}$ in the two loop toplogies.} 
\end{figure} 
from the $2$ and $4$-point functions in $\MSbar$ determines the renormalization
constants given in appendix A. The resulting na\"{\i}ve renormalization group
functions for each operator can be deduced as 
\begin{eqnarray} 
\tilde{\gamma}(g) &=& -~ \frac{(\Nc^2-1)\Nf g^2}{8\Nc\pi^2} ~-~ 
(\Nc^2-1)(2\Nc^2\Nf^2 + \Nc^3 \Nf + 2\Nc^2 + 2) \frac{g^3}{32\Nc^3 \pi^3} \\  
\tilde{\gamma}_m(g) &=& \frac{(\Nc^2-1) g}{2\Nc\pi} ~+~ 
(\Nc^2-1)(2\Nc\Nf - \Nc^2 + 1) \frac{g^2}{8\Nc^2 \pi^2} \nonumber \\  
&& +~ (\Nc^2 - 1)(16\Nf^2\Nc^2 - 20\Nf\Nc^3 + 16\Nc\Nf - 3\Nc^4 + 61\Nc^2 + 62) 
\frac{g^3}{128\Nc^3\pi^3} \\  
\tilde{\beta}(g) &=& \frac{\Nc g^2}{2\pi} ~+~ (4\Nc^3\Nf - 3\Nc^4 + 12\Nc^2 
- 36) \frac{g^3}{16\Nc^2\pi^2} \nonumber \\
&& +~ (10\Nc^4\Nf^2 - 15\Nc^5\Nf + 60\Nc^3\Nf - 180\Nc\Nf+ 87\Nc^4 + 13\Nc^2 
- 360) \frac{g^4}{64\Nc^3\pi^3} \nonumber \\ \\ 
\tilde{\beta}_{30}(g) &=& -~ \frac{(\Nc^2-1)g^2}{8\Nc^2\pi} ~-~ 
(\Nc^2-1)(2\Nc\Nf + 3\Nc^2 - 2) \frac{g^3}{16\Nc^3\pi^2} \nonumber \\  
&& -~ (\Nc^2-1)(96\Nc^2\Nf^2 + 240\Nc^3\Nf - 144\Nc\Nf + 773\Nc^4 - 1004\Nc^2
+ 284) \frac{g^4}{1024\Nc^4\pi^3} \nonumber \\ \\ 
\tilde{\beta}_{31}(g) &=& -~ \frac{(\Nc^2-4)g^2}{8\Nc\pi} ~-~ 
(\Nc^2-4)(\Nc\Nf - \Nc^2 - 1) \frac{g^3}{8\Nc^2\pi^2} \nonumber \\ 
&& -~ (\Nc^2-4)(24\Nc^2\Nf^2 - 20\Nc^3\Nf - 56\Nc\Nf + 45\Nc^4 - 54\Nc^2
+ 94) \frac{g^4}{256\Nc^4\pi^3} \\ 
\tilde{\beta}_{50}(g) &=& -~ \frac{(\Nc^2-1)(\Nc^2 - 4)g^3}{128\Nc^3\pi^2} \\  
\tilde{\beta}_{51}(g) &=& \frac{(\Nc^2-3)g^3}{32\Nc^2\pi^2} \\  
\tilde{\beta}_{10}(g) &=& -~ \frac{3(\Nc^2-1)(\Nc^2 - 4)g^3}{16\Nc^3\pi^2} 
\end{eqnarray} 
where aside from $\OO_{30}$ and $\OO_{31}$ we have only retained those terms of
the evanescent operator $\beta$-functions which are relevant for the true three
loop renormalization group functions. To complete the calculation we need the 
subsequent terms in the projection functions. 

For the $2$-point function this necessitates first of all computing the 
additional graphs to $\la \psi \NN [ \OO_{30} ] \bar{\psi} \ra$ and $\la \psi 
\NN [ \OO_{31} ] \bar{\psi} \ra$ as well as the new Green's functions $\la \psi
\NN [ \OO_{10} ] \bar{\psi} \ra$, $\la \psi \NN [ \OO_{50} ] \bar{\psi} \ra$ 
and $\la \psi \NN [ \OO_{51} ] \bar{\psi} \ra$. As the details of the 
calculation of the latter three are very much akin to those of the former two 
at the previous order we merely record the results  
\begin{eqnarray} 
\rho^{(50)}_m(g) &=& \frac{2}{\pi} ~+~ O(g) ~~~,~~~ \rho^{(51)}_m(g) ~=~ 
\frac{(\Nc^2-1)}{\Nc\pi} ~+~ O(g) \\  
\rho^{(10)}_m(g) &=& \frac{1}{2\pi} \left[ 1 - \ln \left( 
\frac{\tilde{\mu}^2}{m^2} \right) \right] ~+~ O(g) ~.  
\end{eqnarray} 
For $\OO_{3i}$ their composite operator insertion into the $2$-point function
requires the computation of the graphs of figure 7 as well as the contribution 
from the graphs defined by the counterterms to the simple poles in (\ref{o30}) 
and (\ref{o31}) and their projection into the relevant operator. The upshot of 
including the additional graphs is that after $\MSbar$ renormalization we now 
have  
\begin{eqnarray} 
\la \psi \NN [ \OO_{30} ] \bar{\psi} \ra_{wf} &=& -~ \frac{(\Nc^2-1)g} 
{2\Nc\pi^2} ~+~ O(g^2) \nonumber \\  
\la \psi \NN [ \OO_{31} ] \bar{\psi} \ra_{wf} &=& \frac{(\Nc^2-1)g} 
{4\Nc^2\pi^2} ~+~ O(g^2) 
\end{eqnarray} 
and 
\begin{eqnarray} 
\la \psi \NN [ \OO_{30} ] \bar{\psi} \ra_m &=& \frac{1}{\pi} ~-~ 
\frac{7(\Nc^2-1)g}{4\Nc\pi^2} \left[ 2 + \ln \left( \frac{\tilde{\mu}^2}{m^2} 
\right) \right] ~+~ O(g^2) \\ 
\la \psi \NN [ \OO_{31} ] \bar{\psi} \ra_m &=& \frac{(\Nc^2-1)}{2\Nc\pi} ~+~ 
\frac{(\Nc^2-1)g}{\Nc^2\pi^2} \left[ \frac{13}{8}\Nc^2 + \frac{7}{4} - 
\left( \frac{\Nc^2}{2} - \frac{7}{8} \right) \ln \left( 
\frac{\tilde{\mu}^2}{m^2} \right) \right] ~+~ O(g^2) ~. \nonumber 
\end{eqnarray} 
To determine the extra terms of the projection formul{\ae} the insertion of the
right side of (\ref{projdef}) in a $2$-point function must be included. 
Therefore, there is a contribution from the first graph of figure 10 after 
renormalization, from the $\NN[\OO_{11}]$ term of (\ref{projdef}). This enters 
with the coefficients $C^{(30)}(g)$ and $C^{(31)}(g)$. Additionally the second 
graph, being $O(g)$, gives a potential contribution. However, it only 
contributes to the equation for $\rho^{(k)}_m(g)$. Finally, 
\begin{figure}[ht] 
\epsfig{file=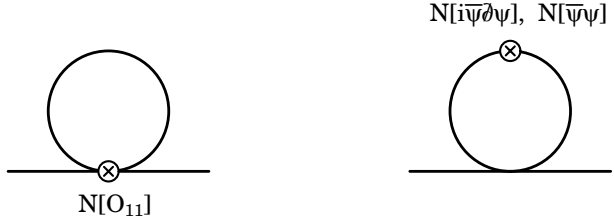,width=10cm} 
\vspace{0.5cm} 
\caption{Graphs for the three loop projection formula in a $2$-point function 
for the NATM.} 
\end{figure} 
the finite part of the right side of (\ref{projdef}) is  
\begin{equation} 
\frac{3(\Nc^2-1)g}{2\Nc\pi^2} \left[ 1 - \ln \left( \frac{\tilde{\mu}^2}{m^2} 
\right) \right] ~-~ \left( \rho^{(30)}(g) + \rho^{(30)}_m(g) \right) \left[ 
1 + \frac{(\Nc^2-1)g}{4\Nc\pi} \left( 3 - \ln \left( 
\frac{\tilde{\mu}^2}{m^2} \right) \right) \right] ~+~ O(g^2) 
\end{equation} 
for $\OO_{30}$ and 
\begin{eqnarray} 
&& \frac{3(\Nc^2-1)(\Nc^2-4)g}{8\Nc^2\pi^2} \left[ 1 - \ln 
\left( \frac{\tilde{\mu}^2}{m^2} \right) \right] \nonumber \\ 
&& -~ \left( \rho^{(31)}(g) + \rho^{(31)}_m(g) \right) 
\left[ 1 + \frac{(\Nc^2-1)g}{4\Nc\pi} 
\left( 3 - \ln \left( \frac{\tilde{\mu}^2}{m^2} \right) \right) \right] ~+~
O(g^2) 
\end{eqnarray} 
for $\OO_{31}$. Therefore, solving this perturbatively we find 
\begin{eqnarray} 
\rho^{(30)}(g) &=& -~ \frac{(\Nc^2-1)g}{2\Nc\pi^2} ~+~ O(g^2) ~~~,~~~ 
\rho^{(31)}(g) ~=~ \frac{(\Nc^2-1)g}{4\Nc^2\pi^2} ~+~ O(g^2) \\  
\rho^{(30)}_m(g) &=& -~ \frac{1}{\pi} ~+~ \frac{25(\Nc^2-1)g}{4\Nc\pi^2} ~+~ 
O(g^2) \\ 
\rho^{(31)}_m(g) &=& -~ \frac{(\Nc^2-1)}{2\Nc\pi} ~+~ 
\frac{(\Nc^2-1)}{\Nc^2} \left[ \frac{3}{4} \ln \left( \frac{\tilde{\mu}^2}{m^2}
\right) - \frac{(7\Nc^2+31)}{8} \right] \frac{g}{\pi^2} ~+~ O(g^2) ~.  
\end{eqnarray} 
\begin{figure}[ht] 
\epsfig{file=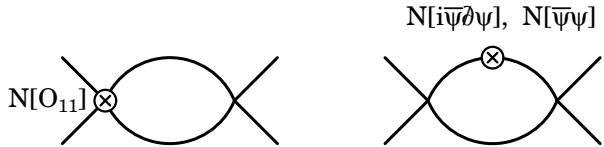,width=10cm} 
\vspace{0.5cm} 
\caption{Graphs for the three loop projection formula in a $4$-point function 
for the NATM.} 
\end{figure} 

The procedure for the $4$-point function is similar. One also has extra graphs
to consider involving the insertion of the relevant operators $\NN[i\bar{\psi} 
\partialslash \psi]$ and $\NN[\bar{\psi} \psi]$ and these are given in 
figure 11 for the right side of (\ref{projdef}). Including counterterm 
contributions both for the evanescent operator poles and the $\OO_{11}$ 
vertex renormalization, the finite two dimensional Green's functions now 
become  
\begin{eqnarray} 
\left. \la \psi \bar{\psi} \NN [ \OO_{30} ] \psi \bar{\psi} \ra 
\right|_{\OO_{11}} &=& -~ \frac{6ig}{\pi} ~+~ i\left[ 3 \Nc \ln \left( 
\frac{\tilde{\mu}^2}{m^2} \right) ~-~ \frac{3\Nf}{2} ~+~ \frac{65\Nc}{4} ~-~ 
\frac{20}{\Nc} \right] \frac{g^2}{\pi^2} ~+~ O(g^3) \nonumber \\ 
\left. \la \psi \bar{\psi} \NN [ \OO_{31} ] \psi \bar{\psi} \ra 
\right|_{\OO_{11}} &=& -~ \frac{3i(\Nc^2-4)g}{2\Nc\pi} ~+~ i\left[ 
\frac{3(\Nc-4)}{4} \ln \left( \frac{\tilde{\mu}^2}{m^2} \right) ~-~ 
\frac{3(\Nc^2-4)\Nf}{8\Nc} \right. \nonumber \\ 
&& \left. \quad\quad\quad\quad\quad\quad\quad\quad ~~-~ 
\frac{(13\Nc^4-29\Nc^2-82)}{4\Nc^2} \right] \frac{g^2}{\pi^2} ~+~ O(g^3) ~. 
\end{eqnarray} 
Likewise, the sum of contributions to the right side give 
\begin{equation} 
[C^{(3i)}(g) + 2g \rho^{(3i)}(g)] \left[ 1 - \left( \frac{\Nc}{2} \ln \left( 
\frac{\tilde{\mu}^2}{m^2} \right) - \frac{3\Nc}{4} \right) \right] ~+~ 
\frac{\Nc\rho^{(3i)}_m(g)g^2}{2\pi} ~+~ O(g^3) 
\end{equation}  
for $i$ $=$ $0$ and $1$. Hence, it is easy to deduce the projection functions
to the order we require are 
\begin{eqnarray} 
C^{(30)}(g) &=& -~ \frac{6g}{\pi} ~-~ \left[ \frac{3\Nf}{2} - \frac{89\Nc}{4}
+ \frac{21}{\Nc} \right] \frac{g^2}{\pi^2} ~+~ O(g^3) \nonumber \\ 
C^{(31)}(g) &=& -~ \frac{3(\Nc^2-4)g}{2\Nc\pi} ~-~ \left[ \frac{3(\Nc^2-4)\Nf}
{8\Nc} + \frac{(15\Nc^4-16\Nc^2-168)}{8\Nc^2} \right] \frac{g^2}{\pi^2} ~+~ 
O(g^3) ~. \nonumber \\ 
\end{eqnarray} 
For the operators $\OO_{10}$, $\OO_{50}$ and $\OO_{51}$ we note that the one
loop graphs of figure 6 with the appropriate operator insertions take simpler
forms. Respectively, we have 
\begin{equation} 
\la \psi \bar{\psi} \NN [ \OO_{10} ] \psi \bar{\psi} \ra ~=~ \frac{ig}{\pi} 
\left[ \frac{1}{\epsilon} \OO_{31} ~+~ 2\OO_{01} ~+~ \frac{1}{2} \ln \left( 
\frac{\tilde{\mu}^2}{m^2} \right) \OO_{31} ~+~ O(\epsilon) \right] ~+~ O(g^2) 
\label{o10} 
\end{equation} 
and 
\begin{eqnarray} 
\la \psi \bar{\psi} \NN [ \OO_{50} ] \psi \bar{\psi} \ra &=& \frac{ig}{\pi} 
\left[ \frac{1}{\epsilon} \left( 40\OO_{31} ~-~ \frac{16(\Nc^2-1)}{\Nc} 
\OO_{50} ~+~ \OO_{71} \right) \right. \nonumber \\ 
&& \left. ~~~~-~ 20 \left( 3 + \ln \left( \frac{\tilde{\mu}^2}{m^2} \right) 
\right) \OO_{31} ~-~ 10 \OO_{41} ~+~ \frac{1}{2} \ln \left( 
\frac{\tilde{\mu}^2}{m^2} \right) \OO_{71} \right. \nonumber \\  
&& \left. ~~~~-~ \frac{2(\Nc^2-1)}{\Nc} \left( 1 + 4 \ln \left( 
\frac{\tilde{\mu}^2}{m^2} \right) \right) \OO_{50} ~+~ O(\epsilon) \right] +~ 
O(g^2) ~.  
\label{o50} 
\end{eqnarray} 
The expression for $\la \psi \bar{\psi} \NN [ \OO_{51} ] \psi \bar{\psi} \ra$
is similar to that for $\OO_{50}$ but involves more operators. As it does not
contain any relevant new features we do not record it. Since the operator
$\OO_{11}$ does not appear on the right side of (\ref{o10}) and (\ref{o50}), 
then after renormalization and removal of the evanescent operators the 
contribution to the left side of the respective projection formul{\ae} is
zero at this order. Therefore we have the simple results  
\begin{equation} 
C^{(10)}(g) ~=~ 0 ~+~ O(g^2) ~~,~~ C^{(50)}(g) ~=~ 0 ~+~ O(g^2) ~~,~~  
C^{(51)}(g) ~=~ 0 ~+~ O(g^2)  
\end{equation} 
which imply that at this order $\tilde{\beta}_{10}(g)$, $\tilde{\beta}_{50}(g)$ 
and $\tilde{\beta}_{51}(g)$ are not in fact needed for the renormalization 
group functions.  With the corrections to the other projection functions, it is
now possible to deduce the true $\MSbar$ renormalization group functions for 
the NATM at three loops. For $SU(\Nc)$ we find 
\begin{eqnarray} 
\beta(g) &=& \mu \frac{\partial g(\mu)}{\partial\mu} ~=~ \frac{\Nc g^2}{2\pi} 
{}~+~ \frac{\Nf\Nc g^3}{4\pi^2} \nonumber \\
&&+~ \left[ \frac{5}{32} \Nc \Nf^2 ~+~ \frac{3}{64}\Nc^3 ~-~ \frac{11}{64}\Nc 
{}~+~ \frac{39}{64} \frac{1}{\Nc} \right] \frac{g^4}{\pi^3} ~+~ O(g^5) 
\end{eqnarray}  
\begin{equation} 
\gamma(g) ~=~ -~ \frac{(\Nc^2-1)\Nf g^2}{8\Nc\pi^2} ~-~ (\Nc^2-1) \left( 2\Nf^2 
+ \Nf \Nc + 1 \right) \frac{g^3}{32\Nc \pi^3} 
\end{equation}  
and 
\begin{eqnarray} 
\gamma_m(g) &=& \frac{(\Nc^2-1) g}{2\Nc\pi} ~+~ (\Nc^2-1)(4\Nf - \Nc) 
\frac{g^2}{16\Nc \pi^2} \nonumber \\  
&& +~ (\Nc^2 - 1) \left( 16\Nf^2\Nc^2 - 12\Nf\Nc^3 + 3\Nc^4 + 5\Nc^2 - 26 
\right) \frac{g^3}{128\Nc^3\pi^3} ~.  
\end{eqnarray}  
It is worth noting that the term involving $\ln(\tilde{\mu}^2/m^2)$ has 
cancelled in the sum for $\gamma_m(g)$. 

Although we have detailed the projection technique of \cite{15} for the massive
NATM there is an alternative method of determining these renormalization group
functions. In performing the calculation we were careful to compute the
Feynman graphs as functions of $d$. Therefore, we can deduce the finite
parts of the $2$ and $4$-point functions after renormalization. These must 
satisfy the full renormalization group equation (\ref{fullrge}) with the 
above functions. Therefore with  
\begin{eqnarray} 
G^{(2)}(p,m,\tilde{\mu},g) &=& i(\pslash - m) ~+~ \frac{(\Nc^2-1)mg}{4\Nc\pi} 
\left[ \ln \left( \frac{\tilde{\mu}^2}{m^2} \right) - 1 \right] \nonumber \\  
&& +~ \pslash \left[ 2\Nc\Nf \ln \left( \frac{\tilde{\mu}^2}{m^2} \right) + 1 
\right] \frac{(\Nc^2-1) g^2}{32\Nc^2\pi^2} \nonumber \\ 
&& -~ m \left[ (2\Nc^2-1) \ln^2 \left( \frac{\tilde{\mu}^2}{m^2} \right) ~-~ 
(7\Nc^2 - 6 + 2\Nc\Nf) \ln \left( \frac{\tilde{\mu}^2}{m^2} \right) \right. 
\nonumber \\ 
&& \left. ~~~~~~~+~ 3\Nc^2 ~-~ 3 \frac{}{} \right] 
\frac{(\Nc^2-1)g^2}{32\Nc^2\pi^2} \nonumber \\ 
&& +~ \pslash \left[ -~ 6\Nc^3\Nf \ln^2 \left( \frac{\tilde{\mu}^2}{m^2} 
\right) ~+~ 6(3\Nc^2 + 2\Nc\Nf - 2)\Nc\Nf \ln \left( \frac{\tilde{\mu}^2}{m^2} 
\right) \right. \nonumber \\ 
&& \left. ~~~~~~-~ 10 \Nc^3\Nf + 4\Nc^2\Nf^2 - 5\Nc^2 + 24\Nc\Nf - 14 \frac{}{} 
\right] \frac{(\Nc^2-1)g^3}{384\Nc^3\pi^3} \nonumber \\ 
&& +~ m \left[ 2(2\Nc^2-1)(3\Nc^2-1) \ln^3 \left( 
\frac{\tilde{\mu}^2}{m^2} \right) \right. \nonumber \\ 
&& \left. ~~~~~~~-~ (84\Nc^4 + 36\Nc^3\Nf - 108\Nc^2 - 12\Nc\Nf + 30) \ln^2 
\left( \frac{\tilde{\mu}^2}{m^2} \right) \right. \nonumber \\ 
&& \left. ~~~~~~~+~ (129\Nc^4 + 60\Nc^3\Nf + 24\Nc^2\Nf^2 - 207\Nc^2 - 84\Nc\Nf 
+ 12 ) \ln \left( \frac{\tilde{\mu}^2}{m^2} \right) \right. \nonumber \\ 
&& \left. ~~~~~~~-~ 45\Nc^4 + 20\Nc^3\Nf + 16\Nc^2\Nf^2 + 42\Nc^2 - 32\Nc\Nf
- 194 \right] \frac{(\Nc^2-1)g^3}{768\Nc^3\pi^3} \nonumber \\  
\end{eqnarray} 
and 
\begin{eqnarray} 
\left. G^{(4)}(0,m,\tilde{\mu},g) \right|_{\OO_{11}} &=& ig ~+~ \frac{\Nc}{8} 
\left[ 3 - 2 \ln \left( \frac{\tilde{\mu}^2}{m^2} \right) \right] 
\frac{g^2}{\pi} \nonumber \\ 
&&+~ \left[ 4 \Nc^4 \ln^2 \left( \frac{\tilde{\mu}^2}{m^2} \right) ~-~ 
(20\Nc^4 - 8\Nc^2 + 8\Nc\Nf ) \ln \left( \frac{\tilde{\mu}^2}{m^2} \right) 
\right. \nonumber \\ 
&& \left. \frac{}{} ~~~~+~ 15\Nc^4 ~+~ 4\Nc^3\Nf + 2\Nc^2 ~-~ 34 \right] 
\frac{g^3}{64\Nc^2\pi^2} \nonumber \\ 
&&-~ \left[ 24 \Nc^3 \ln^3 \left( \frac{\tilde{\mu}^2}{m^2} \right) ~-~ 
12(19\Nc^3 + 2\Nc^2\Nf - 10\Nc + 8\Nf) \ln^2 \left( 
\frac{\tilde{\mu}^2}{m^2} \right) \right. \nonumber \\ 
&& \left. \frac{}{} ~~~~+~ \left( 522\Nc^3 + 192\Nc^2\Nf + 24\Nc\Nf^2
- 456\Nc + 120\Nf \right. \right. \nonumber \\ 
&& \left. \left. ~~~~~~~~~~~ + \frac{96\Nf^2}{\Nc} - \frac{96\Nf}{\Nc^2} 
\right) \ln \left( \frac{\tilde{\mu}^2}{m^2} \right) - 345\Nc^3 - 20\Nc^2\Nf 
- 32\Nc\Nf^2 \right. \nonumber \\ 
&& \left. ~~~~~+~ 764\Nc - 288\Nf + \frac{32\Nf^2}{\Nc} - \frac{1440}{\Nc} + 
\frac{1056\Nf}{\Nc} + \frac{1424}{\Nc^3} \right] \frac{g^4}{1536\pi^3} ~.  
\nonumber \\ 
\label{4ptfin} 
\end{eqnarray}  
it is easy to verify that (\ref{fullrge}) holds. If we did not possess the
$\MSbar$ functions then we could have solved for them by ensuring that the
renormalization group equation (\ref{fullrge}) holds. For the massless model, 
since the mass is not a parameter of the theory it means that this avenue is 
not possible for us. This will leave us only the projection method which we 
have now shown to be consistent. In writing (\ref{4ptfin}) we have retained 
only the contribution to the original relevant operator since after 
renormalization $d$ is set to $2$ and the evanescent operators removed. 
Finally, it is a simple task to rewrite the $\MSbar$ $\beta$-function back in 
terms of the colour group Casimirs as  
\begin{eqnarray} 
\beta(\lambda) &=& -~ \frac{\CG \lambda^2}{2\pi} ~+~ \frac{T(R)\Nf\CG 
\lambda^3}{2\pi^2} \nonumber \\
&&-~ \CG \left[ \frac{5}{8} T^2(R) \Nf^2 ~+~ \frac{39}{16} C^2_2(R) ~-~ 
\frac{67}{32}\CR\CG ~+~ \frac{31}{64} C^2_2(G) \right] \frac{\lambda^4}{\pi^3} 
{}~+~ O(\lambda^5) \nonumber \\ 
\end{eqnarray}  
where we have also included the coupling $\lambda$ for comparison with the two 
loop results of \cite{14}. 

\sect{Massless NATM in $\MSbar$.} 

As the massive model clearly does not provide us with a fully chirally 
symmetric theory, we have also computed the $\MSbar$ renormalization constants
for the model when the propagator (\ref{masslessprop}) is used. As the 
calculation runs very close to that of the previous section we concentrate on 
the new features. First, when one evaluates the contributing graphs to three
loops with (\ref{masslessprop}) the structure is simpler than the corresponding
ones obtained by using (\ref{massiveprop}). Indeed the first graph of figure 4 
now becomes  
\begin{eqnarray} 
i g^4 I^3 \left[ \frac{}{} \right. &-& \left. 
(105d^3 - 420d^2 + 588d - 272)(\Nc^2 - 4)(\Nc^2 - 1) 
\frac{\OO_{10}}{128\Nc^2} \right. \nonumber \\ 
&-& \left. \left( 4(2\Nc^2\Nf^2 - 4(d-2)\Nc\Nf + 3(d-2)^2)(d-2)^3\Nf\Nc ~-~ 
4(d-2)^6 \right. \right. \nonumber \\ 
&& ~~+ \left. \left. (105d^3 - 420d^2 + 588d - 272)(\Nc^4 - 4\Nc^2 + 10)\Nc^2
\right) \frac{\OO_{11}}{64\Nc^3} \right. \nonumber \\ 
&+& \left. 7(15d^2 - 70d + 88) \! \left( \! (\Nc^4 - 2\Nc^2 + 6)(\Nc^2 - 1)  
\frac{\OO_{30}}{128\Nc^4} ~+~ (\Nc^4 + 2)(\Nc^2 - 4) \frac{\OO_{31}}{64\Nc^3} 
\right) \right. \nonumber \\ 
&-& \left. 7(3d - 10) \left( (\Nc^2 - 4)(\Nc^2 - 1)  
\frac{\OO_{50}}{128\Nc^2} ~+~ (\Nc^4 - 4\Nc^2 + 10) \frac{\OO_{51}}{64\Nc} 
\right) \right. \nonumber \\ 
&+& \left. (\Nc^4 - 2\Nc^2 + 6)(\Nc^2 - 1)  
\frac{\OO_{70}}{128\Nc^4} ~+~ (\Nc^4 + 2)(\Nc^2 - 4) \frac{\OO_{71}}{64\Nc^3} 
{}~~ \right] ~.  
\end{eqnarray}  
Comparing this with the expression for the same graph in the massive case we 
observe its decomposition into the operator basis involves only those operators
with $\Gamma_{(2r+1)}$ for integer $r$. This is a result of the chiral 
symmetry which underlies the choice (\ref{masslessprop}). Moreover, this 
is also a property of all the other graphs to this order, so that the potential
need for renormalization constants $Z_{ni}$ for $n$ even and $i$ $=$ $0$ or
$1$ is automatically excluded. Completing the $\MSbar$ renormalization of
the model as previously it turns out that the renormalization constants are
the same as for the massive model with exception of $Z_m$ which, of course,
does not arise at the outset.  

The computation of the full $\gamma(g)$ and $\beta(g)$ in the massless NATM now
follows the use of the projection formula for the massive model whose practical
use we have just demonstrated. Since $m$ is no longer a true parameter of the 
model, the projection formula (\ref{projdef}) clearly takes the simpler form,
\cite{15},  
\begin{equation} 
\int d^d x \, \NN [ \OO_{ki} ] ~=~ \int d^d x \left. \left( \, \rho^{(ki)}(g) 
\NN [ i \bar{\psi} \partialslash \psi ~+~ 2g \OO_{11} ] ~+~ C^{(ki)}(g) 
\NN [ \OO_{11} ] \right) \right|_{ g_i = 0 ~ d = 2 } ~. 
\label{projdefmassless} 
\end{equation}  
The absence of $\rho^{(ki)}_m(g)$ and, in principle, different values for
the contributing terms to the Green's function structure might mean that the
true $\MSbar$ renormalization group functions will differ from those of the
massive model. However, as argued in \cite{15,22} one would expect them to be 
equivalent. Therefore, it is important that this is checked explicitly. 
Though we need only do this for $\beta(g)$, as clearly $\gamma(g)$ will be
unchanged. 

For the $4$-point function there are again no contributions to the relevant
operator in the Green's function involving separate insertions of $\OO_{10}$,
$\OO_{50}$ and $\OO_{51}$ leaving us to concentrate on $\OO_{30}$ and 
$\OO_{31}$. For these, in order to illustrate the differences in the structure
of the Green's functions which are needed to deduce $C^{(30)}(g)$ and 
$C^{(31)}(g)$, we note that their one loop forms are 
\begin{eqnarray} 
\la \psi \bar{\psi} \NN [ \OO_{30} ] \psi \bar{\psi} \ra &=& \frac{ig}{\pi} 
\left[ \frac{1}{\epsilon} \left( \OO_{51} ~-~ \frac{(\Nc^2-1)}{\Nc} \OO_{30} 
\right) ~-~ 6\OO_{11} \right. \nonumber \\ 
&& \left. ~~~~+~ \epsilon \left( 4 - 3 \ln \left( \frac{\tilde{\mu}^2}{m^2} 
\right) \right) \OO_{11} ~+~ O(\epsilon;\OO_{k0},\OO_{k1}) \right] ~+~ O(g^2) 
\end{eqnarray} 
and 
\begin{eqnarray} 
\la \psi \bar{\psi} \NN [ \OO_{31} ] \psi \bar{\psi} \ra &=& \frac{ig}{\pi} 
\left[ \frac{1}{\epsilon} \left( \frac{(\Nc^2+4)}{\Nc} \OO_{31} ~+~ 
\frac{(\Nc^2-1)}{4\Nc^2} \OO_{50} ~+~ \frac{(\Nc^2-4)}{4\Nc} \OO_{51} \right) 
\right. \nonumber \\ 
&& \left. ~~~~-~ \frac{3(\Nc^2-4)}{2\Nc} \OO_{11} ~+~ 
O(\epsilon;\OO_{k0},\OO_{k1}) \right] ~+~ O(g^2) 
\end{eqnarray} 
which can be compared with those of the massive NATM calculation. Further, 
including all relevant counterterms the restriction of these Green's functions 
in $d$ $=$ $2$ at two loops gives  
\begin{eqnarray} 
\left. \la \psi \bar{\psi} \NN [ \OO_{30} ] \psi \bar{\psi} \ra 
\right|_{\OO_{11}} &=& -~ \frac{6ig}{\pi} ~+~ i\left[ 3 \Nc \ln \left( 
\frac{\tilde{\mu}^2}{m^2} \right) ~-~ \frac{9\Nf}{2} ~+~ \frac{67\Nc}{4} ~-~ 
\frac{20}{\Nc} \right] \frac{g^2}{\pi^2} ~+~ O(g^3) \nonumber \\ 
\left. \la \psi \bar{\psi} \NN [ \OO_{31} ] \psi \bar{\psi} \ra 
\right|_{\OO_{11}} &=& -~ \frac{3i(\Nc^2-4)g}{2\Nc\pi} ~+~ i\left[ 
\frac{3(\Nc-4)}{4} \ln \left( \frac{\tilde{\mu}^2}{m^2} \right) ~-~ 
\frac{9(\Nc^2-4)\Nf}{8\Nc} \right. \nonumber \\ 
&& \left. \quad\quad\quad\quad\quad\quad\quad\quad ~~-~ 
\frac{(6\Nc^4-14\Nc^2-41)}{2\Nc^2} \right] \frac{g^2}{\pi^2} ~+~ O(g^3)
\end{eqnarray} 
which differ from those of the massive model. To deduce the final values of
the projection functions, we record that now the right side of 
(\ref{projdefmassless}) is simpler than (\ref{projdef}) due to the absence of
$\rho^{(k)}_m(g)$. Explicitly, we have  
\begin{equation} 
[C^{(3i)}(g) + 2g \rho^{(3i)}(g)] \left[ 1 - \left( \frac{\Nc}{2} \ln \left( 
\frac{\tilde{\mu}^2}{m^2} \right) - \frac{\Nf}{2} - \frac{3\Nc}{4} \right) 
\right] ~+~ O(g^3) 
\end{equation}  
for $i$ $=$ $0$ and $1$. However, when solving perturbatively for the
projection functions it turns out that their values to this order are the
same as before. Therefore the $\MSbar$ $\beta$-function is identical to that
of the massive model as expected. The discrepancy between each set of 
formul{\ae} rests in the hidden role of the mass term in various Green's
functions. 
\vspace{0.5cm} 
\begin{figure}[hb] 
\epsfig{file=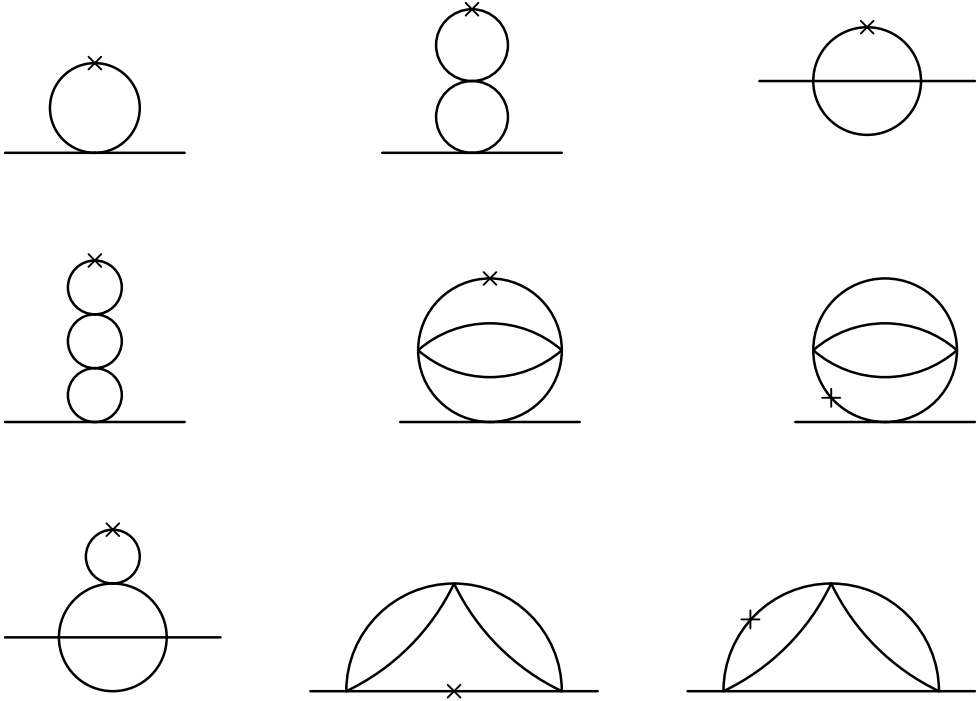,width=12.9cm} 
\vspace{0.5cm} 
\caption{Graphs contributing to the renormalization of $[\bar{\psi}\psi]$ in
the massless NATM.} 
\end{figure} 

Although it may seem that for the massless model that there is no 
renormalization group function for the mass, it is possible to compute the 
anomalous dimension of the composite operator whose coupling would correspond 
to a mass. Clearly this is $[\bar{\psi}\psi]$. As this operator will be 
renormalized, one can define an anomalous dimension 
$\gamma_{\bar{\psi}\psi}(g)$ from its renormalization constant $Z_{\bar{\psi}
\psi}$ which can then be regarded as the mass dimension. Indeed the computation
of the three loop quark mass dimension in $\MSbar$ in QCD was performed in the 
massless theory by renormalizing the $2$-point quark Green's function with a 
$[\bar{\psi}\psi]$ insertion, \cite{13}. We have computed the relevant diagrams
to three loops for both the massless Gross Neveu model, in order to check our 
integration routines, and the massless NATM. These are illustrated in figure 12
where we have only shown those diagrams which are not trivially zero. Also, the
cross indicates the location of the insertion of $[\bar{\psi}\psi]$. Therefore,
if the insertion in the two loop tower of bubbles had not been on the top loop 
but the lower one, then the diagram is simply zero because in the massless 
model the one loop tadpole vanishes by Lorentz symmetry. Therefore if a 
topology has one tadpole then for a non-zero contribution its 
$[\bar{\psi}\psi]$ insertion must be in that tadpole. Although it may appear 
that we have to introduce a new set of integrals to carry out the calculation, 
we have exploited the fact that since we are only interested in the 
$[\bar{\psi}\psi]$ renormalization, the graphs need only be computed at zero 
external momentum. In this case it turns out that tensor integrals for each 
graph can be paired with a graph used in the evaluation of the $4$-point 
function. The topology of the related graph is deduced by regarding the 
insertion as the location of two external legs. So, for example, the last graph
of figure 12 is paired with the penultimate graph of figure 4. Similarly the 
first graph of the bottom row of figure 12 is related to the fifth graph of 
figure 4. This observation ensures that we do not have to unnecessarily 
undertake extra calculation. It turns out that performing the $\MSbar$ 
renormalization of these graphs that $Z_{\bar{\psi} \psi}$ is identical to 
$Z_m$ of the massive model for both the Gross Neveu model and the NATM. To 
complete the evaluation of the associated true renormalization group function, 
we need to extend the projection formula of (\ref{projdefmassless}) to the case
of bona fide operator insertions. In this case we take it to be 
\begin{eqnarray} 
\int d^d x \, \NN [ \OO_k ] &=& \int d^d x \left. \left( \frac{}{} \, 
\rho^{(k)}(g) \NN [ i \bar{\psi} \partialslash \psi ] ~+~ \left( C^{(k)}(g) 
+ 2g\rho^{(k)}(g) \right) \NN [ \OO_{11}] \right. \right. \nonumber \\ 
&& \left. \left. ~~~~~~~~~+~ \left( \rho^{(k)}(g) + 
\rho^{(k)}_{\bar{\psi}\psi}(g) \right) \NN [1] \frac{}{} \right) 
\right|_{ g_i = 0 ~ d = 2 } ~.  
\end{eqnarray}  
Here the last term has the insertion of unity since we regard this operator
equation as being inserted in a Green's function of the structure 
$\la \psi ~.~ [\bar{\psi} \psi] \bar{\psi} \ra$ which includes the original
operator 
\begin{figure}[hb] 
\epsfig{file=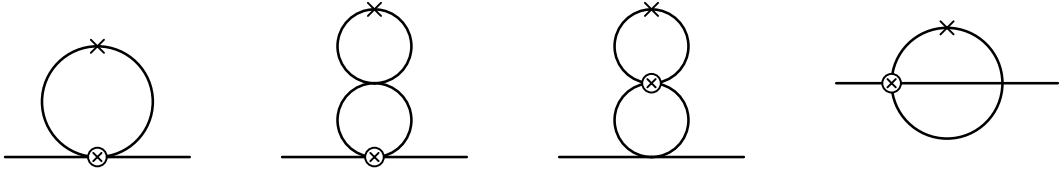,width=14cm} 
\vspace{0.5cm} 
\caption{Graphs contributing to the massless NATM $2$-point function with 
$[\bar{\psi}\psi]$ insertion from the evanescent operators $\OO_{3i}$.} 
\end{figure} 
insertion itself represented by the dot. Therefore, one needs to be 
careful in distinguishing which operator is which when considering the graphs 
contributing to the Green's function. To this end we have illustrated the 
relevant non-zero  contributions to the two loop Green's function $\la \psi 
\NN[\OO] [\bar{\psi} \psi] \bar{\psi} \ra$ in figure 13. As we have detailed 
similar calculations earlier, we record the values of the Green's function for 
the operators $\NN[\OO_{30}]$ and $\NN[\OO_{31}]$ as those of other operators 
will only involve one loop graphs. We found, 
\begin{eqnarray} 
\la \psi \NN [ \OO_{30} ] [\bar{\psi}\psi] \bar{\psi} \ra &=& -\, 
\frac{1}{\pi} ~-~ \frac{(\Nc^2-1)g}{\Nc\pi^2} \left[ \frac{9}{4} + \frac{7}{4} 
\ln \left( \frac{\tilde{\mu}^2}{m^2} \right) \right] ~+~ O(g^2) \nonumber \\  
\la \psi \NN [ \OO_{31} ] [\bar{\psi}\psi] \bar{\psi} \ra &=& -\,  
\frac{(\Nc^2-1)}{2\Nc\pi} ~-~ \frac{(\Nc^2-1)g}{\Nc^2\pi^2} \! \left[ 
\frac{15\Nc^2}{8} + \frac{9}{8} - \left( \frac{\Nc^2}{2} - \frac{7}{8} \right) 
\ln \! \left( \frac{\tilde{\mu}^2}{m^2} \right) \right] \,+~ O(g^2) 
\nonumber \\ 
\end{eqnarray} 
where we note that the overall sign difference between these and the analogous
results in the massive model is due to the fact that here we are inserting
the operator $[\bar{\psi}\psi]$ and not the negative of this. Moreover, 
\begin{eqnarray} 
\rho^{(50)}_{\bar{\psi}\psi}(g) &=& \frac{2}{\pi} ~+~ O(g) ~~~,~~~ 
\rho^{(51)}_{\bar{\psi}\psi}(g) ~=~ \frac{(\Nc^2-1)}{\Nc\pi} ~+~ O(g) \\  
\rho^{(10)}_{\bar{\psi}\psi}(g) &=& \frac{1}{2\pi} \left[ 2 - \ln \left( 
\frac{\tilde{\mu}^2}{m^2} \right) \right] ~+~ O(g)  
\end{eqnarray} 
so that the numerical values of $\rho^{(10)}_{\bar{\psi}\psi}(g)$ and 
$\rho^{(10)}_m(g)$ differ. For the right side of the projection formula we have
displayed the one loop corrections which need to be computed in figure 14. With
their values the respective right sides of each projection formul{\ae} are  
\begin{equation} 
-~ \frac{3(\Nc^2-1)g}{2\Nc\pi^2} \! \left[ 2 - \ln \left( 
\frac{\tilde{\mu}^2}{m^2} \right) \right] ~-~ \left( \rho^{(30)}(g) 
+ \rho^{(30)}_{\bar{\psi}\psi}(g) \right) \!\! \left[ 1 + 
\frac{(\Nc^2-1)g}{4\Nc\pi} \left( 2 - \ln \left( \frac{\tilde{\mu}^2}{m^2} 
\right) \! \right) \right] ~+~ O(g^2) 
\end{equation} 
for $\OO_{30}$ and 
\begin{eqnarray} 
&& -~ \frac{3(\Nc^2-1)(\Nc^2-4)g}{8\Nc^2\pi^2} \left[ 2 - \ln \left( 
\frac{\tilde{\mu}^2}{m^2} \right) \right] \nonumber \\ 
&& -~ \left( \rho^{(31)}(g) + \rho^{(31)}_{\bar{\psi}\psi}(g) \right) 
\left[ 1 + \frac{(\Nc^2-1)g}{4\Nc\pi} 
\left( 2 - \ln \left( \frac{\tilde{\mu}^2}{m^2} \right) \right) \right] ~+~ 
O(g^2) 
\end{eqnarray} 
for $\OO_{31}$, from which it is easy to deduce 
\begin{eqnarray} 
\rho^{(30)}_{\bar{\psi}\psi}(g) &=& -~ \frac{1}{\pi} ~+~ 
\frac{25(\Nc^2-1)g}{4\Nc\pi^2} ~+~ O(g^2) \nonumber \\ 
\rho^{(31)}_{\bar{\psi}\psi}(g) &=& -~ \frac{(\Nc^2-1)}{2\Nc\pi} ~+~ 
\frac{(\Nc^2-1)}{\Nc^2} \left[ \frac{3}{4} \ln \left( \frac{\tilde{\mu}^2}{m^2}
\right) - \frac{(7\Nc^2+37)}{8} \right] \frac{g}{\pi^2} ~+~ O(g^2) ~.  
\end{eqnarray} 
Therefore, the $\MSbar$ $\gamma_{\bar{\psi}\psi}(g)$ is identical to 
$\gamma_m(g)$ of the massive model.  
\vspace{0.5cm} 
\begin{figure}[ht]  
\epsfig{file=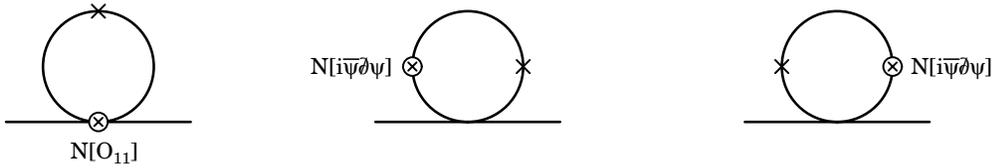,width=14cm} 
\vspace{0.5cm} 
\caption{Graphs for the three loop projection formula in a $2$-point function
with $[\bar{\psi}\psi]$ insertion for the massless NATM.} 
\end{figure} 

\sect{Symmetric schemes.} 

Although we have produced the renormalization group functions for both massive
and massless versions of the NATM in $\MSbar$ there still remains the problem
of reconciling the results with the equivalences with other models discussed
earlier. The $\MSbar$ expressions do not represent results which are consistent
with symmetries of the original model and the problem can effectively be traced
to a hidden $\gamma^5$ issue, \cite{15,21,22}. In other words, when one deals 
with models with continuous chiral symmetries in dimensional regularization 
this symmetry is not preserved in $\MSbar$. For example, in QCD one must treat 
$\gamma^5$ carefully in renormalizing quark bilinear currents involving 
$\gamma^5$. There the resolution is to use a renormalization scheme which does 
preserve the symmetry. This is achieved by first performing an $\MSbar$ 
calculation to the requisite order and then introduce an additional finite 
renormalization to ensure that in that case the Green's functions obey the 
(quantum) Ward identities, including the axial anomaly, \cite{39}. For the 
present calculation the general approach is the same. Though there will be 
several minor differences. One rests in the fact that the analogous finite 
renormalization is chosen so that one ensures the renormalization group 
functions in the first instance and not the Green's functions are the same in 
the equivalences between the various models. Further, in the present context, 
similar finite renormalization schemes have been established for the three loop
anomalous dimension ensuring agreement of the equivalences between the ABTM and
the Gross Neveu model for $\Nf$ $=$ $1$, \cite{22}. In essence that scheme was 
specified by removing in addition to the simple poles, the (numerical) finite 
part of the Green's function into the renormalization constant, \cite{22}. We 
stress that terms involving $\ln(\tilde{\mu}^2/m^2)$ were not removed so that 
one is still within the context of mass independent renormalization schemes. In
\cite{15,22} this was referred to as the symmetric scheme. To appreciate the 
effect such a choice of scheme can have on the renormalization group functions 
we have determined them for the Gross Neveu model. To three loops for the two
dimensional massive model, we found that 
\begin{eqnarray} 
\beta(g) &=& -~ \frac{(\Nf-1)g^2}{\pi} ~+~ \frac{(\Nf-1)g^3}{2\pi^2} ~-~ 
\frac{(6\Nf-1)(\Nf-1)g^4}{8\pi^3} ~+~ O(g^5) \\ 
\gamma(g) &=& -~ \frac{(2\Nf-1)g^2}{8\pi^2} ~-~ 
\frac{(2\Nf-1)(2\Nf-3)g^3}{8\pi^3} ~+~ O(g^4) \\  
\gamma_m(g) &=& -~ \frac{(2\Nf-1)g}{2\pi} ~-~ 
\frac{(2\Nf-1)(4\Nf-7)g^2}{8\pi^2} \nonumber \\ 
&& -~ \frac{(8\Nf^2-42\Nf+37)(2\Nf-1)g^3}{16\pi^3} ~+~ O(g^4) ~.  
\label{gnrgefin} 
\end{eqnarray} 
Clearly, one now has a cubic polynomial in $\Nf$ at three loops in 
$\gamma_m(g)$ in contrast to the quadratic that appears in the $\MSbar$ scheme.
We can readily produce this result since we have been careful in the 
contruction of the integration routines to keep all the $d$-dependence. The aim
now is to extend our calculation to determine the scheme or schemes which 
preserve the known equivalences of the NATM. We follow the approach of 
\cite{15,22} and reconcile the results with the $\MSbar$ forms of the $O(2\Nf)$
Gross Neveu model which possesses a discrete $\gamma^5$ chiral symmetry. Though
we stress that one could choose another scheme as reference. Further, since the
choice of scheme is very much arbitrary we will take a general approach and
introduce a parametrization. This will involve a set of variables which will be 
constrained by the equivalences. To illustrate this we will consider the 
massless model only and define the finite terms of the wave function 
renormalization to be 
\begin{equation}  
\frac{(\Nc^2-1)g^2}{\Nc\pi^2} \left( w_{21}\Nf + w_{22}\Nc + \frac{w_{23}}{\Nc}
\right) ~.  
\label{Zgen1}
\end{equation} 
Likewise for the coupling constant renormalization constant, $Z_{11}Z^2_\psi$,
we take the finite part to be 
\begin{equation}  
\left( b_{11}\Nf + b_{12}\Nc + \frac{b_{13}}{\Nc} \right) \frac{g}{\pi} ~+~ 
\left( b_{21}\Nf^2 + b_{22}\Nf\Nc + \frac{b_{23}\Nf}{\Nc} + b_{24}\Nc^2
+ b_{25} + \frac{b_{26}}{\Nc^2} \right) \frac{g^2}{\pi^2} ~.  
\label{Zgen2}
\end{equation}  
An $O(g^3)$ term will affect the four loop form of the renormalization group 
functions. The form of the additional terms has been chosen to be consistent 
with the colour Casimir structure that is allowed. Therefore, the term of 
(\ref{Zgen1}) is proportional to $C_2(R)$. Of course, these additional terms 
will affect the form of poles in $\epsilon$ of the renormalization constants 
which we have not recorded. 

In the context of the usual multiplicatively renormalizable theories the 
treatment of a scheme change is straightforward and well documented. (See, 
for example, \cite{40,21,22}.) Once one has one scheme, the renormalization
group functions in another are determined by the relations 
\begin{eqnarray}  
\bar{\beta}(\bar{g}) &=& \beta(g) \frac{\partial ~}{\partial g} \bar{g}(g) 
\nonumber \\ 
\bar{\gamma}(\bar{g}) &=& \gamma(g) ~-~ 2 \beta(g) 
\frac{\partial ~}{\partial g} \ln \xi(g)  
\label{rgegen} 
\end{eqnarray}  
where 
\begin{equation} 
\bar{G}^{(n)}(p,\mu,\bar{g}) ~=~ \xi^n G^{(n)}(p,\mu,g) 
\label{gfdefn} 
\end{equation} 
and the bar denotes the corresponding quantity in another scheme. In general 
the relation between the couplings in both schemes is determined by comparing 
the renormalized $4$-point Green's functions using (\ref{gfdefn}) after the 
function $\xi(g)$ is deduced at the appropriate order in the $2$-point 
function. For the non-multiplicatively renormalizable NATM the situation is the
same, though of course one has to first compute the new Green's functions in 
the scheme involving (\ref{Zgen1}) and (\ref{Zgen2}). To understand fully the 
proper procedure for this we have first performed the renormalization in the 
massive model and then determined the renormalization group functions by 
finding the function $\xi(g)$ before applying the formul{\ae}, (\ref{rgegen}). 
These could then be checked against the renormalization group functions deduced
by ensuring that the renormalization group equation itself was satisfied. 
Repeating this first method for the massless model we have the general 
functions 
\begin{eqnarray} 
\xi(g) &=& \left[ 1 ~+~ \frac{(\Nc^2-1)g^2}{\Nc\pi^2} \left( w_{21}\Nf ~+~ 
w_{22}\Nc ~+~ \frac{w_{23}}{\Nc} \right) \right] \nonumber \\  
\bar{g}(g) &=& g \left[ 1 ~-~ \left( b_{11}\Nf ~+~ b_{12}\Nc ~+~ 
\frac{b_{13}}{\Nc} \right) \frac{g}{\pi} ~+~ \left( 2\Nc^2\Nf^2 b_{11}^2 ~-~ 
\Nc^2\Nf^2 b_{21} \right. \right. \nonumber \\ 
&& \left. \left. ~~+~ 2\Nc^3\Nf w_{21} ~+~ 4\Nc^3\Nf b_{11}b_{12} ~-~ 
\Nc^3\Nf b_{22} ~-~ 2\Nc\Nf w_{21} ~+~ 4\Nc\Nf b_{11} b_{13} \right. \right.
\nonumber \\ 
&& \left. \left. ~~-~ \Nc\Nf b_{23} ~+~ 2\Nc^4 w_{22} ~+~ 
2\Nc^4 b_{12}^2 ~-~ \Nc^4 b_{24} ~-~ 2\Nc^2 w_{22} ~+~ 2\Nc^2 w_{23} 
\right. \right. \nonumber \\ 
&& \left. \left. ~~+~ 4\Nc^2 b_{12} b_{13} ~-~ \Nc^2 b_{25} ~-~ 2w_{23} ~+~ 
2b_{13}^2 ~-~ b_{26} \right) \frac{g^2}{\Nc^2\pi^2} \right]  
\end{eqnarray} 
which therefore provide us with the general scheme results 
\begin{eqnarray} 
\bar{\gamma}(\bar{g}) &=& -~ \frac{(\Nc^2 - 1)\Nf}{8\Nc} 
\frac{\bar{g}^2}{\pi^2} ~-~ \frac{(\Nc^2 - 1)}{32\Nc^2} 
\left[ 8\Nc\Nf^2 b_{11} ~+~ 2\Nc\Nf^2 ~+~ 32\Nc^2\Nf w_{21} ~+~ 
8\Nc^2\Nf b_{12} \right. \nonumber \\ 
&& \left. +~ \Nc^2\Nf ~+~ 8\Nf b_{13} ~+~ 32\Nc^3 w_{22} ~+~ 32\Nc w_{23} ~+~ 
\Nc \right] \frac{\bar{g}^3}{\pi^3} ~+~ O(\bar{g}^4)  
\end{eqnarray} 
and 
\begin{eqnarray} 
\bar{\beta}(\bar{g}) &=& \frac{\Nc \bar{g}^2}{2\pi} ~+~ 
\frac{\Nf\Nc \bar{g}^3}{4\pi^2} ~+~ \left[ 32\Nc^2\Nf^2 b_{11}^2 ~+~ 
16\Nc^2\Nf^2 b_{11} ~-~ 32\Nc^2\Nf^2 b_{21} ~+~ 10\Nc^2\Nf^2 \right. 
\nonumber \\ 
&& \left. +~ 64\Nc^3\Nf w_{21} ~+~ 64\Nc^3\Nf b_{11} b_{12} ~+~ 
16\Nc^3\Nf b_{12} ~-~ 32\Nc^3\Nf b_{22} ~-~ 64\Nc\Nf w_{21} \right. 
\nonumber \\ 
&& \left. +~ 64\Nc\Nf b_{11} b_{13} ~+~ 16\Nc\Nf b_{13} ~-~ 32\Nc\Nf b_{23} ~+~
64\Nc^4 w_{22} ~+~ 32\Nc^4 b_{12}^2 ~-~ 32\Nc^4 b_{24} \right. \nonumber \\
&& \left. +~ 3\Nc^4 ~-~ 64\Nc^2 w_{22} ~+~ 64\Nc^2 w_{23} ~+~ 
64\Nc^2 b_{12} b_{13} ~-~ 32\Nc^2 b_{25} ~-~ 11\Nc^2 ~-~ 64 w_{23} \right. 
\nonumber \\ 
&& \left. +~ 32 b_{13}^2 ~-~ 32 b_{26} ~+~ 39 \right] 
\frac{\bar{g}^4}{64\Nc\pi^3} ~+~ O(\bar{g}^5) ~.  
\end{eqnarray}  
The $\MSbar$ results are of course recovered by setting the various constants
to zero. 

The two equivalences we wish to preserve are those discussed in section 2. For 
the one involving the ABTM and the Gross Neveu model at $\Nf$ $=$ $1$, we first
need to convert the $SU(\Nc)$ dependence into the general colour group Casimirs
and then apply (\ref{abellim}). We find  
\begin{eqnarray} 
\bar{\gamma}(\bar{g}) &=& -~ \frac{\Nf\bar{g}^2}{2\pi^2} ~-~ 
\left( \Nf^2 ~+~ 4b_{11}\Nf^2 - 4b_{13}\Nf \right) \frac{\bar{g}^3}{2\pi^3} 
\nonumber \\ 
\bar{\beta}(\bar{g}) &=& 0 
\end{eqnarray} 
giving 
\begin{eqnarray} 
0 &=& 4b_{11} ~-~ 4b_{13} ~+~ 1 \nonumber \\ 
0 &=& 32w_{21} ~-~ 32w_{23} ~-~ 1 \nonumber \\ 
0 &=& 32b_{21} ~-~ 32b_{23} ~+~ 32b_{26} ~-~ 31  
\end{eqnarray} 
where the latter two equations arise from choosing to remove the numerical 
part of the $2$-point and $4$-point functions respectively at three loops, 
similar to \cite{22}. With the caveat we had previously on the form of 
$\gamma(g)$ for the second equivalence, we find that the constraints in this 
case are 
\begin{eqnarray} 
0 &=& 16b_{12} ~+~ 5b_{13} ~+~ 80w_{21} ~+~ 256w_{22} \nonumber \\ 
0 &=& 8192b_{12}^2 ~+~ 5120b_{12}b_{13} ~+~ 800b_{13}^2 ~-~ 2048b_{22} ~-~ 
640b_{23} ~-~ 8192b_{24} ~-~ 512b_{25} \nonumber \\
&& +~ 480b_{26} ~+~ 4800w_{21} ~+~ 15360w_{22} ~+~ 231 ~.  
\end{eqnarray} 
Clearly not all the general coefficients can be determined by these 
restrictions and the undetermined coefficients represent the remaining freedom 
in the choice of scheme that is possible. For instance, if we set 
\begin{equation} 
w_{21} ~=~ b_{12} ~=~ b_{13} ~=~ 0 ~~,~~ b_{22} ~=~ \frac{15}{128} ~~,~~ 
b_{23} ~=~ \frac{1}{4} ~~,~~ b_{24} ~=~ \frac{3}{64} ~~,~~ b_{26} ~=~ 1 
\label{choice} 
\end{equation} 
then the renormalization group functions will take the form 
\begin{eqnarray} 
\bar{\gamma}(\bar{g}) &=& -~ \frac{(\Nc^2 - 1)\Nf}{8\Nc} 
\frac{\bar{g}^2}{\pi^2} ~-~ \frac{(\Nc^2 - 1)\Nf\bar{g}^3}{32} ~+~ 
O(\bar{g}^4) \nonumber \\  
\bar{\beta}(\bar{g}) &=& \frac{\Nc \bar{g}^2}{2\pi} ~+~ 
\frac{\Nf\Nc \bar{g}^3}{4\pi^2} \nonumber \\ 
&& +~ \left[ 16\Nc^2\Nf^2 - 60\Nc^3\Nf - 128\Nc\Nf + 24\Nc^4 - 105\Nc^2 + 144 
\right] \frac{\bar{g}^4}{1024\Nc\pi^3} ~+~ O(\bar{g}^5) ~. \nonumber \\  
\end{eqnarray} 
In the abelian limit, we obtain $\bar{\beta}(\bar{g})$ $=$ $0$ and 
$\bar{\gamma}(\bar{g})$ $=$ $-$ $\Nf\bar{g}^2/(2\pi^2)$ $+$ $O(\bar{g}^4)$ 
consistent with (\ref{gnrge}). Moreover, comparing $\bar{\gamma}(\bar{g})$ with
the choice made in \cite{22} one observes that there are other schemes aside 
from the symmetric one of \cite{15,22} for which the general form can be 
reconciled. Of course, choices other than (\ref{choice}) can be made. Finally, 
this completes our analysis of the basic renormalization group functions for 
the original massless non-abelian Thirring model.  

\sect{Discussion.} 

We conclude by stressing that we have now completed the first three loop 
renormalization of the NATM in a dimensional regularization. Whilst it involved
the careful extraction and treatment of evanescent operators in the four point
function we were able to determine the renormalization group functions of the 
model at a new order in perturbation theory in a variety of mass independent 
renormalization schemes. Moreover, from a calculational point of view we have 
also demonstrated how one interprets and applies the projection formul{\ae} of  
\cite{15,29} beyond the leading perturbative order which was used to determine
the $2$-loop $\beta$-functions of \cite{15}. Although we have concentrated on 
the renormalization of the NATM, another important property we have established
is the breakdown of multiplicative renormalizability in the Gross Neveu model
at three loops in dimensional regularization. For the NATM this, of course, 
occurs at one loop, \cite{15}. In \cite{22} the consequences of the 
perturbative non-multiplicative renormalizability of the Gross Neveu model were
discussed from the point of view of the (non-perturbative) large $\Nf$ 
expansion and the determination of critical exponents at a fixed point of the 
$d$-dimensional $\beta$-function. Indeed it is worth reviewing some of these 
results now that the property has been established. Such exponents are 
fundamental, being scheme independent, and relate to experimentally measurable 
quantities. One ordinarily determines them from the $d$-dimensional 
renormalization group functions through the $\epsilon$-expansion, using perhaps
resummation techniques to improve the series convergence for applications in, 
say, three dimensions. The large $\Nf$ expansion complements such an approach 
in that perturbation theory is reordered so that bubble chain graphs are summed
first. Indeed in the context of the Gross Neveu model the large $\Nf$ expansion
has been successful in $d$-dimensions in determining the wave function exponent
$\eta$ to $O(1/\Nf^3)$, \cite{41,42,43}, and the mass and $\beta$-function 
exponents to $O(1/\Nf^2)$, \cite{44,45}. For the three loop perturbative 
$\MSbar$ results which had been computed there is exact agreement in the region
of overlap since the exponents are related to the renormalization group 
functions through, for example, $\eta$ $=$ $\gamma(g^\star)$ where $g^\star$ is
the value of the critical coupling in $d$-dimensions. However, in light of the 
perturbative non-multiplicative renormalizability of the Gross Neveu model, it 
is worth first of all recording the basis of the large $\Nf$ formulation. 
Rather than using the Lagrangian (\ref{gnlag}) the starting point is the 
massless Lagrangian written in terms of an auxiliary field, $\sigma$,  
\begin{equation} 
L^{\mbox{\footnotesize{gn}}} ~=~ i \bar{\psi}^{i} \partialslash \psi^{i} ~+~ 
\sigma \bar{\psi}^{i} \psi^{i} ~-~ \frac{\sigma^2}{2g} ~. 
\label{gnlagsig} 
\end{equation}  
One then develops the large $\Nf$ formalism using (\ref{gnlagsig}). With this
form it does not appear possible to generate another operator of the form
$\sigma^{\mu\nu\rho} \bar{\psi}^i \Gamma^{(3)}_{\mu\nu\rho} \psi^i$ which would
be analogous to the evanescent operator generated in (\ref{gnevlag}). For 
instance, if one considers the graphs of figure 4 which give rise to the
evanescent operator, it is evident that a $\Gamma_{(3)}$ $\otimes$ 
$\Gamma_{(3)}$ structure cannot emerge in the $1/\Nf$ expansion. With this 
apparent discrepancy between the renormalized perturbative theory and large
$\Nf$ approach, it has been suggested in \cite{22} that the simple relation
$\eta$ $=$ $\gamma(g^\star)$ needs to be modified in the perturbative approach
to account for the presence of the evanescent operators in $d$-dimensions. In
other words in perturbation theory one assumes that in $d$-dimensions the Gross
Neveu model possesses an infinite set of couplings $g_\kappa$ for each operator
$\half (\bar{\psi} \Gamma_{(\kappa)} \psi)^2$ corresponding to a Lagrangian 
which is multiplicatively renormalizable. Then the fixed point of the 
$d$-dimensional theory is defined by $\beta_{(\kappa)}(g^\star_\rho)$ $=$ $0$ 
for all $\kappa$. Whilst in practical terms this may be difficult to determine,
if we assume that there is a solution then one can define the actual exponents 
by extending the na\"{\i}ve relation for the wave function to $\eta$ $=$ 
$\gamma(g^\star_\rho)$, \cite{22}, where now the renormalization group 
functions will depend on all the couplings. Turning away from this ideal 
scenario one can approach this infinite set of fixed points in an appropriate 
way, \cite{22}, by assuming that it is close to the one defined by $g^\star_0$ 
$\neq$ $0$ and $g^\star_n$ $=$ $0$. In other words the case where all the 
evanescent couplings are set to zero, \cite{22}. Several issues immediately 
arise. First, it is not clear whether the na\"{\i}ve $\beta$-functions of the 
$d$-dimensional theory are the ones which define the fixed points. Second, the 
practical task of computing the corrections to the evanescent fixed points 
needs to be resolved. In reviewing these issues we believe the simplest 
strategy to tackle this problem is to perform explicit calculations. As the 
evanescent operators in the Gross Neveu model do not become evident before 
three loops, the first stage would be to determine the four loop $\MSbar$ 
renormalization group functions. Second, since the renormalization constant for
the evanescent operator is $\Nf$ independent its effect in the $1/\Nf$ 
expansion will not become apparent at least before $O(1/\Nf^3)$. In fact by 
computing the graphs of figure 6 it turns out that they are $\Nf$ independent. 
This follows from the fact that to obtain any $\Nf$ dependence one must have a 
closed fermion loop. However, from the nature of the new evanescent vertex such
a loop in any of the graphs is associated with $\mbox{tr}\,\Gamma^{(3)}$ which 
vanishes. So for the $\beta$-function and the mass dimension one can not access
the contribution from this operator in large $\Nf$ prior to $O(1/\Nf^4)$. 
Therefore to reconcile higher order large $\Nf$ results for critical exponents 
with explicit perturbative calculations will require a huge amount of 
calculation. 

\vspace{0.8cm} 
\noindent 
{\bf Acknowledgements.} We are grateful to Prof. A.N. Vasil'ev for drawing our
attention to the problem of the multiplicative renormalizability of the Gross
Neveu model in dimensional regularization. This work was carried out with the 
support of PPARC through a Postgraduate Studentship (JFB) and an Advanced 
Fellowship (JAG). Invaluable to the calculation were the symbolic manipulation 
programme {\sc Form}, \cite{37}, and computer algebra package {\sc Reduce}, 
\cite{46}. The figures were prepared using the package {\sc FeynDiagram}.  

\appendix 

\sect{$\MSbar$ renormalization constants.}  

In this appendix we record the values for all the renormalization constants in
the $\MSbar$ scheme to three loops in the non-abelian Thirring model. From the
$2$-point function, we find  
\begin{eqnarray} 
Z_\psi &=& 1 ~-~ \frac{(\Nc^2-1)\Nf}{16\Nc\epsilon} \frac{g^2}{\pi^2} ~-~ 
(\Nc^2-1) \left( \frac{\Nf}{48\epsilon^2} ~+~ 
\frac{(2\Nc^2\Nf^2 + \Nc^3\Nf + 2\Nc^2 + 2)}{96\Nc^3\epsilon} \right) 
\frac{g^3}{\pi^3} \nonumber \\ \\  
Z_m &=& 1 ~+~ \frac{(\Nc^2-1)}{2\Nc\epsilon} \frac{g}{\pi} ~+~  
(\Nc^2-1) \left( \frac{(2\Nc^2-1)}{8\Nc^2\epsilon^2} ~+~ 
\frac{(2\Nc\Nf - \Nc^2 + 1)}{16\Nc^2\epsilon} \right) \frac{g^2}{\pi^2}  
\nonumber \\ 
&& +~ (\Nc^2-1) \left( \frac{(3\Nc^2-1)(2\Nc^2-1)}{48\Nc^3\epsilon^3} ~+~
\frac{(14\Nc^3\Nf - 6\Nc\Nf - 6\Nc^4 + 4\Nc^2 - 19)}{96\Nc^3\epsilon^2} \right. 
\nonumber \\
&& \left. ~~~~~~~~~~~~~~~~~+~ \frac{(16\Nc^2\Nf^2 - 20\Nc^3\Nf + 16\Nc\Nf 
- 3\Nc^4 + 61\Nc^2 + 62)}{384\Nc^3\epsilon} \right) \frac{g^3}{\pi^3} ~.  
\end{eqnarray} 
For the $4$-point operators we have 
\begin{eqnarray} 
Z_{11} &=& 1 ~+~ \frac{\Nc}{2\epsilon} \frac{g}{\pi} ~+~ 
\left( \frac{\Nc^2}{4\epsilon^2} ~+~ \frac{(4\Nc^3\Nf - 3\Nc^4 + 12\Nc^2 
- 36)}{32\Nc^2\epsilon} \right) \frac{g^2}{\pi^2} \nonumber \\  
&& +~ \left( \frac{\Nc^3}{8\epsilon^3} ~+~
\frac{(28\Nc^5\Nf - 9\Nc^6 - 72\Nc^4 - 72\Nc^2 + 288)}{192\Nc^3\epsilon^2} 
\right. \nonumber \\
&& \left. ~~~~~~+~ \frac{(10\Nc^4\Nf^2 - 15\Nc^5\Nf + 60\Nc^3\Nf 
- 180\Nc\Nf+ 87\Nc^4 + 13\Nc^2 - 360)}{192\Nc^3\epsilon} \right) 
\frac{g^3}{\pi^3} \nonumber \\ \\  
Z_{30} &=& -~ (\Nc^2-1) \left[ \frac{1}{8\Nc^2\epsilon} \frac{g}{\pi} ~+~ 
\left( \frac{(5\Nc^2-4)}{16\Nc^3\epsilon^2} ~+~ 
\frac{(2\Nc\Nf + 3\Nc^2 - 2)}{32\Nc^3\epsilon} \right) \frac{g^2}{\pi^2} 
\right. \nonumber \\  
&& \left. +~ \left( \frac{(169\Nc^4 - 12\Nc^3\Nf + 12\Nc\Nf - 380\Nc^2 + 388)}
{192\Nc^4\epsilon^3} \right. \right. \nonumber \\
&& \left. \left. ~~~~~~~+~ \frac{(100\Nc^3\Nf - 80\Nc\Nf + 315\Nc^4 - 728\Nc^2 
+ 852)}{384\Nc^4\epsilon^2} \right. \right. \nonumber \\
&& \left. \left. ~~~~~~~+~ \frac{(96\Nc^2\Nf^2 + 240\Nc^3\Nf - 144\Nc\Nf 
+ 773\Nc^4 - 1004\Nc^2 + 284)}{3072\Nc^4\epsilon} \right) 
\frac{g^3}{\pi^3} \right] \\ 
Z_{31} &=& -~ (\Nc^2-4) \left[ \frac{1}{8\Nc\epsilon} \frac{g}{\pi} ~+~ 
\left( -~ \frac{1}{4\Nc^2\epsilon^2} ~+~ 
\frac{(\Nc\Nf - \Nc^2 - 1)}{16\Nc^2\epsilon} \right) \frac{g^2}{\pi^2} 
\right. \nonumber \\  
&& \left. +~ \left( \frac{(7\Nc^4 - 2\Nc^3\Nf + 8\Nc\Nf + 16\Nc^2 + 104)}
{64\Nc^3\epsilon^3} \right. \right. \nonumber \\
&& \left. \left. ~~~~~~+~ \frac{( - 160\Nc\Nf + 105\Nc^4 + 200\Nc^2 
+ 1328)}{768\Nc^3\epsilon^2} \right. \right. \nonumber \\
&& \left. \left. ~~~~~~+~ \frac{(24\Nc^2\Nf^2 - 20\Nc^3\Nf - 56\Nc\Nf 
+ 45\Nc^4 - 54\Nc^2 + 94)}{768\Nc^3\epsilon} \right) 
\frac{g^3}{\pi^3} \right] \\ 
Z_{10} &=& -~ (\Nc^2-1)(\Nc^2-4) \left[ \frac{3}{32\Nc^3\epsilon} 
\frac{g^2}{\pi^2} ~+~ \!\! \left( \frac{(\Nc^2 - 4)}{32\Nc^4\epsilon^2} ~+~ 
\frac{(20\Nc\Nf - 7\Nc^2 + 40)}{256\Nc^4\epsilon} \right) 
\frac{g^3}{\pi^3} \right] \\ 
Z_{50} &=& (\Nc^2-1)(\Nc^2-4) \left[ \left( \frac{1}{64\Nc^3\epsilon^2} ~-~ 
\frac{1}{256\Nc^3\epsilon} \right) \frac{g^2}{\pi^2} \right. \nonumber \\  
&& \left. +~ \left( \frac{(17\Nc^2 - 20)}{192\Nc^4\epsilon^3} ~+~ 
\frac{(20\Nc\Nf - 97\Nc^2 + 104)}{1536\Nc^4\epsilon^2} ~-~ 
\frac{(16\Nc\Nf + 39\Nc^2 + 64)}{3072\Nc^4\epsilon} \right) \frac{g^3}{\pi^3} 
\right] \nonumber \\ \\ 
Z_{51} &=& \left[ \left( \frac{(\Nc^4 - 4\Nc^2 + 12)}{64\Nc^2\epsilon^2} ~+~ 
\frac{(\Nc^2 - 3)}{64\Nc^2\epsilon} \right) \frac{g^2}{\pi^2} 
\right. \nonumber \\  
&& \left. +~ \left( -~ \frac{(7\Nc^6 - 40\Nc^4 + 56\Nc^2 + 160)}
{128\Nc^3\epsilon^3} \right. \right. \nonumber \\
&& \left. \left. ~~~~~~+~ \frac{(5\Nc^5\Nf - 20\Nc^3\Nf + 60\Nc\Nf - 9\Nc^6
+ 19\Nc^4 - 135\Nc^2 + 312)}{384\Nc^3\epsilon^2} \right. \right. \nonumber \\
&& \left. \left. ~~~~~~-~ \frac{(\Nc^5\Nf - 16\Nc^3\Nf + 48\Nc\Nf 
- 71\Nc^4 + 39\Nc^2 + 192)}{768\Nc^3\epsilon} \right) 
\frac{g^3}{\pi^3} \right] \\ 
Z_{70} &=& (\Nc^2-1) \left[ -~ \frac{(\Nc^4 - 4\Nc^2 + 12)}{512\Nc^4\epsilon^3} 
{}~+~ \frac{(3\Nc^4 - 28\Nc^2 + 84)}{6144\Nc^4\epsilon^2} ~+~ 
\frac{(\Nc^2 + 6)(\Nc^2 - 2)}{4096\Nc^4\epsilon} \right] \frac{g^3}{\pi^3} 
\nonumber \\ \\ 
Z_{71} &=& -~ (\Nc^2-4) \left[ \frac{(\Nc^4 + 8)}{512\Nc^3\epsilon^3} ~-~ 
\frac{7}{768\Nc^3\epsilon^2} ~+~ \frac{1}{512\Nc^3\epsilon} \right] 
\frac{g^3}{\pi^3} ~.  
\end{eqnarray}  
For completeness, we also provide the values for the same renormalization
constants in the abelian Thirring model in the $\MSbar$-scheme. These were
computed separately from those of the NATM, but agree with them when the
abelian limit of (\ref{abellim}) is taken. We have 
\begin{eqnarray} 
Z_\psi &=& 1 ~-~ \frac{\Nf}{4\epsilon} \frac{g^2}{\pi^2} ~-~ 
\frac{(\Nf^2 + 1)}{6\epsilon} \frac{g^3}{\pi^3} \\ 
Z_m &=& 1 ~+~ \frac{1}{\epsilon} \frac{g}{\pi} ~+~ \left( \frac{1}{2\epsilon^2}
{} ~+~ \frac{(2\Nf - 1)}{4\epsilon} \right) \frac{g^2}{\pi^2} \nonumber \\ 
&& +~ \left( \frac{1}{6\epsilon^3} ~+~ \frac{(6\Nf - 19)}{12\epsilon^2} 
{} ~+~ \frac{(8\Nf^2 - 8\Nf + 31)}{24\epsilon} \right) \frac{g^3}{\pi^3} \\ 
Z_{10} &=& 1 ~-~ \frac{3}{2\epsilon} \frac{g^2}{\pi^2} ~-~ \left( \frac{4} 
{\epsilon^2} ~+~ \frac{5(\Nf - 2)}{2\epsilon} \right) \frac{g^3}{\pi^3} \\ 
Z_{30} &=& -~ \left[ \frac{1}{2\epsilon} \frac{g}{\pi} ~+~ \left( 
\frac{2}{\epsilon^2} ~+~ \frac{(\Nf + 1)}{2\epsilon} \right) \frac{g^2}{\pi^2} 
\right. \nonumber \\  
&& \left. ~~~~~+~ \left( -~ \frac{(3\Nf - 59)}{3\epsilon^3} ~+~ 
\frac{(20\Nf + 119)}{6\epsilon^2} ~+~ \frac{(8\Nf^2 + 12\Nf + 39)}{16\epsilon} 
\right) \frac{g^3}{\pi^3} \right] \\ 
Z_{50} &=& \left[ \left( \frac{1}{4\epsilon^2} ~-~ \frac{1}{16\epsilon} \right)
\frac{g^2}{\pi^2} ~+~ \left( \frac{10}{3\epsilon^3} ~+~ 
\frac{(5\Nf - 26)}{12\epsilon^2} ~-~ \frac{(\Nf - 4)}{6\epsilon} \right) 
\frac{g^3}{\pi^3} \right] \\ 
Z_{70} &=& -~ \left[ \frac{1}{8\epsilon^3} ~-~ \frac{7}{96\epsilon^2} ~+~ 
\frac{1}{64\epsilon} \right] \frac{g^3}{\pi^3} ~. 
\end{eqnarray}  
From these renormalization constants, it is straightforward to compute the 
na\"{\i}ve renormalization group functions to $O(g^3)$. They also agree with 
the abelian limit of the analogous functions in the NATM, apart from 
$\tilde{\beta}_{10}(g)$. We find  
\begin{eqnarray} 
\tilde{\gamma}(g) &=& -~ \frac{\Nf g^2}{2\pi^2} ~-~ 
\frac{(\Nf^2 + 1)g^3}{2\pi^3} \\  
\tilde{\gamma}_m(g) &=& \frac{g}{\pi} ~+~ \frac{(2\Nf - 1)g^2}{2\pi^2} ~+~ 
\frac{(8\Nf^2 - 8\Nf + 31)g^3}{8\pi^3} \\  
\tilde{\beta}(g) &=& -~ \frac{3g^3}{\pi^2} ~-~ \frac{15(\Nf - 2)g^4}{2\pi^4} \\ 
\tilde{\beta}_{30}(g) &=& -~ \frac{g^2}{2\pi} ~-~ \frac{(\Nf + 1)g^3}{\pi^2} \\ 
\tilde{\beta}_{50}(g) &=& -~ \frac{g^3}{8\pi^2} ~.  
\end{eqnarray} 
Finally, we note that the values of the $\MSbar$ renormalization group 
functions to three loops are 
\begin{eqnarray} 
\beta(g) &=& 0 \nonumber \\ 
\gamma(g) &=& -~ \frac{\Nf g^2}{2\pi^2} ~-~ \frac{\Nf^2g^3}{2\pi^3} 
\nonumber \\ 
\gamma_m(g) &=& \frac{g}{\pi} ~+~ \frac{\Nf g^2}{\pi^2} ~+~ 
\frac{(8\Nf^2-13)g^3}{8\pi^3} ~.  
\end{eqnarray}


\newpage
\begin{thebibliography}{99} 
\bibitem{1} D. Gross \& A. Neveu, Phys. Rev. {\bf D10} (1974), 3235. 
\bibitem{2} R. Dashen \& Y. Frishman, Phys. Lett. {\bf B46} (1973), 439; Phys.
Rev. {\bf D11} (1975), 2781. 
\bibitem{3} A. Hasenfratz \& P. Hasenfratz, Phys. Lett. {\bf B297} 166. 
\bibitem{4} J.A. Gracey, Phys. Lett. {\bf B318} (1993), 177. 
\bibitem{5} J.A. Gracey, Phys. Lett. {\bf B373} (1996), 178. 
\bibitem{6} J.A. Gracey, Phys. Lett. {\bf B322} (1994), 141; J.F. Bennett \&
J.A. Gracey, Nucl. Phys. {\bf B517} (1998), 241.  
\bibitem{7} M. Ciuchini, S.\'{E}. Derkachov, J.A. Gracey \& A.N. Manashov,
hep-ph/9903410. 
\bibitem{8} T. van Ritbergen, J.A.M. Vermaseren \& S.A. Larin, Phys. Lett. 
{\bf B400} (1997), 379. 
\bibitem{9} J.A.M. Vermaseren, S.A. Larin \& T. van Ritbergen, Phys. Lett. 
{\bf B405} (1997) 327; K.G. Chetyrkin, Phys. Lett. {\bf B404} (1997) 161.  
\bibitem{10} D.J. Gross \& F.J. Wilczek, Phys. Rev. Lett. {\bf 30} (1973), 1343;
H.D. Politzer, Phys. Rev. Lett. {\bf 30} (1973), 1346.
\bibitem{11} W.E. Caswell, Phys. Rev. Lett. {\bf 33} (1974), 244; D.R.T. Jones,
Nucl. Phys. {\bf B75} (1974), 531; E.S. Egorian \& O.V. Tarasov, Teor. Mat. 
Fiz. {\bf 41} (1979), 26.  
\bibitem{12} O.V. Tarasov, A.A. Vladimirov \& A.Yu. Zharkov, Phys. Lett. 
{\bf 93B} (1980), 429; S.A. Larin \& J.A.M. Vermaseren, Phys. Lett. {\bf B303} 
(1993), 334.
\bibitem{13} D.V. Nanopoulos \& D.A. Ross, Nucl. Phys. {\bf B157} (1979) 273; 
R. Tarrach, Nucl. Phys. {\bf B183} (1981) 384; O. Nachtmann \& W. Wetzel, Nucl.
Phys. {\bf B187} (1981) 333; O. Tarasov, JINR preprint P2-82-900.  
\bibitem{14} C. Destri, Phys. Lett. {\bf B210} (1988), 173; Phys. Lett. 
{\bf B213} (1988), 565(E). 
\bibitem{15} A. Bondi, G. Curci, G. Paffuti \& P. Rossi, Ann. Phys. {\bf 199} 
(1990), 268. 
\bibitem{16} W. Wetzel, Phys. Lett. {\bf B153} (1985), 297. 
\bibitem{17} J.A. Gracey, Nucl. Phys. {\bf B341} (1990), 403. 
\bibitem{18} J.A. Gracey, Nucl. Phys. {\bf B367} (1991), 657. 
\bibitem{19} C. Luperini \& P. Rossi, Ann. Phys. {\bf 212} (1991), 371.  
\bibitem{20} A. Bondi, G. Curci, G. Paffuti \& P. Rossi, Phys. Lett. {\bf B216} 
(1989), 349. 
\bibitem{21} A.N. Vasil'ev, M.I. Vyazovskii, S.\'{E}. Derkachov \& N.A. Kivel, 
Theor. Math. Phys. {\bf 107} (1996), 27. 
\bibitem{22} A.N. Vasil'ev, M.I. Vyazovskii, S.\'{E}. Derkachov \& N.A. Kivel, 
Theor. Math. Phys. {\bf 107} (1996), 359. 
\bibitem{23} S.J. Brodsky \& P. Huet, Phys. Lett. {\bf B417} (1998), 145. 
\bibitem{24} C.R. Hagen, Nuovo Cim. {\bf 51B} (1967), 169; Nuovo Cim. {\bf 51A}
(1967), 1033; A.H. Mueller \& T.L. Trueman, Phys. Rev. {\bf D4} (1971), 1635;
Y. Taguchi, A. Tanaka \& K. Yamamoto, Prog. Theor. Phys. {\bf 52} (1974), 1042; 
S. Hikami \& T. Muta, Prog. Theor. Phys. {\bf 57} (1977), 785. 
\bibitem{25} D. Kutasov, Phys. Lett. {\bf B227} (1989), 68. 
\bibitem{26} A.D. Kennedy, J. Math. Phys. {\bf 22} (1981), 1330.  
\bibitem{27} A.N. Vasil'ev, S.\'{E}. Derkachov \& N.A. Kivel, Theor. Math. Phys.
{\bf 103} (1995), 179. 
\bibitem{28} G. Curci \& G. Paffuti, Nucl. Phys. {\bf B286} (1987), 399.  
\bibitem{29} W. Zimmermann, Ann. Phys. {\bf 77} (1973), 536; Ann. Phys. 
{\bf 77} (1973), 570. 
\bibitem{30} M. Bos, Phys. Lett. {\bf B189} (1987), 435; Ann. Phys. {\bf 181} 
(1988), 177. 
\bibitem{31} Z-M. Xi, Phys. Lett. {\bf B214} (1988), 204; Nucl. Phys. 
{\bf B314} (1989), 112.  
\bibitem{32} M.J. Dugan \& B. Grinstein, Phys. Lett. {\bf B256} (1991), 239.  
\bibitem{33} S. Herrlich \& U. Nierste, Nucl. Phys. {\bf B455} (1995), 39; N. 
Pott, hep-ph/9710503.  
\bibitem{34} T. van Ritbergen, A.N. Schellekens \& J.A.M. Vermaseren, Int. J.
Mod. Phys. {\bf A14} (1999), 41.  
\bibitem{35} P. Cvitanovic, Phys. Rev. {\bf D14} (1976), 1536. 
\bibitem{36} P. Nogueira, J. Comput. Phys. {\bf 105} (1993), 406.  
\bibitem{37} J.A.M. Vermaseren, {\it Form} version 2.2c, (CAN publication,  
Amsterdam, 1992). 
\bibitem{38} N.A. Kivel, A.S. Stepanenko \& A.N. Vasil'ev, Nucl. Phys. {\bf 
B424} (1994), 619. 
\bibitem{39} T.L. Trueman, Phys. Lett. {\bf B88} (1979), 331; S.A. Larin, Phys.
Lett. {\bf B303} (1993), 113.  
\bibitem{40} J.C. Collins, {\it Renormalization} (Cambridge University Press, 
1984). 
\bibitem{41} J.A. Gracey, Int. J. Mod. Phys. {\bf A6} (1991), 395, 2755(E).  
\bibitem{42} A.N. Vasil'ev, S.\'{E}. Derkachov, N.A. Kivel \& A.S. Stepanenko,
Theor. Math. Phys. {\bf 94} (1993), 179. 
\bibitem{43} J.A. Gracey, Int. J. Mod. Phys. {\bf A9} (1994), 727.  
\bibitem{44} A.N. Vasil'ev, \& A.S. Stepanenko, Theor. Math. Phys. {\bf 97} 
(1993), 364.  
\bibitem{45} J.A. Gracey, Phys. Lett. {\bf B297} (1992), 293; Int. J. Mod. 
Phys. {\bf A9} (1994), 567.   
\bibitem{46} A.C. Hearn, {\it Reduce Users Manual} version 3.4, (Rand 
publication CP78, 1991). 
\end{thebibliography}
\end{document}